\documentclass[journal=jctcce, manuscript=article]{achemso}

\usepackage[table]{xcolor}
\usepackage[version=3]{mhchem}
\usepackage{amsfonts}
\usepackage{amsmath}
\usepackage{amsthm}
\usepackage{amssymb}
\usepackage{graphicx}
\graphicspath{{img/}}
 \DeclareGraphicsExtensions{.pdf} %png for low res. pdf for higg res. I also have svg as well but other packages needed
\usepackage{caption}
\usepackage{subcaption}
 \DeclareCaptionSubType*[arabic]{figure}
\usepackage{braket}
\usepackage{multicol}
\usepackage{todonotes}
\usepackage{booktabs}
\usepackage[nohyperlinks,nolist]{acronym}
\usepackage{algorithm}
\usepackage{algpseudocode}
\usepackage{siunitx}
\usepackage{mathtools}
\usepackage{hyperref}
\usepackage[inline]{enumitem}
\usepackage{svg}

\newlength{\dch}
\newlength{\sch}
\newcommand{\ud}{\ensuremath{\,\mathrm{d}}}

\newcommand{\scaling}{\phi}

\newcommand{\newton}{\ensuremath{V_{\rho}}}
\newcommand{\pot}{\ensuremath{V}}

\newcommand{\potsol}{\ensuremath{V_{\textrm{R}}}}
\newcommand{\permittivity}{\ensuremath{\varepsilon(\vect{r})}}
\newcommand{\rhoeff}{\ensuremath{\rho_{\textrm{eff}}}}

\newcommand{\polarization}{\ensuremath{\gamma}}

\newcommand*{\diff}{\mathop{}\!\mathrm{d}} % Differential

\newcommand*{\pderiv}[3][]{%
\frac{\partial^{#1}#2}% Partial derivative
{\partial #3^{#1}}}
\newcommand*{\vect}[1]{\boldsymbol{#1}} % Vectorx'
\newcommand*{\mat}[1]{\boldsymbol{#1}}% Matrix

\newcommand{\opdm}[2]{\rho_{1}\left(#1, #2\right)}
\newcommand{\rdm}{\opdm{\vect{r}}{\vect{r}^{\prime}}}
\newcommand{\dirac}[2]{\delta\left(#1 - #2\right)}
\newcommand{\density}[1]{\rho\left(#1\right)}
\newcommand{\dens}{\density{\vect{r}}}
\DeclareMathOperator\erf{erf}

\newcommand{\major}{\cellcolor{blue!100}}
\newcommand{\minor}{\cellcolor{blue!25}}

\author{Gabriel A. Gerez S.}
\affiliation{Hylleraas Centre for Quantum Molecular Sciences, Department of Chemistry, UiT The Arctic University of Norway, N-9037 Troms{\o}, Norway}

\author{Roberto Di Remigio Eikås}
\affiliation{Algorithmiq Ltd, Kanavakatu 3C, FI-00160 Helsinki, Finland}
\email{roberto@algorithmiq.fi}

\author{Stig Rune Jensen}
\affiliation{Hylleraas Centre for Quantum Molecular Sciences, Department of Chemistry, UiT The Arctic University of Norway, N-9037 Troms{\o}, Norway}

\author{Magnar Bjørgve}
\affiliation{Hylleraas Centre for Quantum Molecular Sciences, Department of Chemistry, UiT The Arctic University of Norway, N-9037 Troms{\o}, Norway}

\author{Luca Frediani}
\affiliation{Hylleraas Centre for Quantum Molecular Sciences, Department of Chemistry, UiT The Arctic University of Norway, N-9037 Troms{\o}, Norway}
\email{luca.frediani@uit.no}

\title{Cavity-free continuum solvation: implementation and parametrization in a multiwavelet framework}

\SectionNumbersOn
\begin{document}

\begin{acronym}
\acro{AO}{atomic orbital}
\acro{API}{Application Programmer Interface}
\acro{AUS}{Advanced User Support}
\acro{BEM}{Boundary Element Method}
\acro{BO}{Born-Oppenheimer}  
\acro{CBS}{complete basis set}
\acro{CC}{Coupled Cluster}
\acro{CTCC}{Centre for Theoretical and Computational Chemistry}
\acro{CoE}{Centre of Excellence}
\acro{DC}{dielectric continuum}  
\acro{DFT}{density functional theory}  
\acro{DKH}{Douglas-Kroll-Hess}
\acro{EFP}{effective fragment potential}
\acro{ECP}{effective core potential}
\acro{EU}{European Union}
\acro{GGA}{generalized gradient approximation}
\acro{GPE}{Generalized Poisson Equation}
\acro{GTO}{Gaussian Type Orbital}
\acro{HF}{Hartree-Fock}  
\acro{HPC}{high-performance computing}
\acro{Hylleraas}[HC]{Hylleraas Centre for Quantum Molecular Sciences}
\acro{IEF}{Integral Equation Formalism}
\acro{IGLO}{individual gauge for localized orbitals}
\acro{KB}{kinetic balance}
\acro{KS}{Kohn-Sham}
\acro{LAO}{London atomic orbital}
\acro{LAPW}{linearized augmented plane wave}
\acro{LDA}{local density approximation}
\acro{MAD}{mean absolute deviation}
\acro{maxAD}{maximum absolute deviation}
\acro{MM}{molecular mechanics}  
\acro{MCSCF}{multiconfiguration self consistent field}
\acro{MPA}{multiphoton absorption}
\acro{MRA}{multiresolution analysis}
\acro{MSDD}{Minnesota Solvent Descriptor Database}
\acro{MW}{multiwavelet}
\acro{NAO}{numerical atomic orbital}
\acro{NeIC}{nordic e-infrastructure collaboration}
\acro{KAIN}{Krylov-accelerated inexact Newton}
\acro{NMR}{nuclear magnetic resonance}
\acro{NP}{nanoparticle}  
\acro{OLED}{organic light emitting diode}
\acro{PAW}{projector augmented wave}
\acro{PBC}{Periodic Boundary Condition}
\acro{PCM}{polarizable continuum model}
\acro{PW}{plane wave}
\acro{QC}{quantum chemistry}  
\acro{QM/MM}{quantum mechanics/molecular mechanics}  
\acro{QM}{quantum mechanics}  
\acro{RCN}{Research Council of Norway}
\acro{RMSD}{root mean square deviation}
\acro{RKB}{restricted kinetic balance}
\acro{SC}{semiconductor}
\acro{SCF}{self-consistent field}
\acro{STSM}{short-term scientific mission}
\acro{SAPT}{symmetry-adapted perturbation theory}
\acro{SERS}{surface-enhanced raman scattering}
\acro{WPREL}[WP1]{Work Package 1}
\acro{WPROP}[WP2]{Work Package 2}
\acro{WPAPP}[WP3]{Work Package 3}
\acro{WP}{Work Package}  
\acro{X2C}{exact two-component}
\acro{ZORA}{zero-order relativistic approximation}
\acro{ae}{almost everywhere}
\acro{BVP}{boundary value problem}
\acro{PDE}{partial differential equation}
\acro{RDM}{1-body reduced density matrix}
\acro{SCRF}{self-consistent reaction field}
\acro{IEFPCM}{Integral Equation Formalism \ac{PCM}}
\acro{FMM}{fast multipole method}
\acro{DD}{domain decomposition}
\end{acronym}

%  \acro{SCF}{Self consistent field}
%  \acro{PCM}{Polarizable Continuum Model}
%  \acro{MW}{Multiwavelet}
%  \acro{PE}{Poisson Equation}

\maketitle  
\begin{abstract}
We present a multiwavelet-based implementation of a quantum/classical polarizable continuum model.
The solvent model uses a diffuse solute-solvent boundary and a position-dependent permittivity, lifting 
the sharp-boundary assumption underlying many existing continuum solvation models.
We are able to include both surface and volume polarization effects in the quantum/classical coupling, 
with guaranteed precision, due to the adaptive refinement strategies of our multiwavelet implementation.
The model can account for complex solvent environments and does not need 
\emph{a posteriori} corrections for volume polarization effects. 
We validate our results against a sharp-boundary continuum model 
and find very good correlation of the polarization energies computed for the Minnesota solvation database.
\end{abstract}

\section{Introduction}
Continuum solvation models have been used in quantum chemistry for
half a
century\cite{Miertus:1981fr,Tomasi:2005ipa,Tomasi:1994wt,Cramer:1999wt}. Their use
is motivated by the need to simulate the effect of a large solvent
environment on a molecular solute, keeping at the same time the
computational cost to a minimum.

Several models and flavors have throughout the years been
developed. Common to essentially all such models are two basic
assumptions: 
\begin{enumerate*}[label={\arabic*)}]
    \item the solvent degrees of freedom can be conveniently
described in terms of a continuum, parameterized using macroscopic properties of the solvent;
\item the quantum system is confined inside a cavity and the
solute-solvent interaction is described in terms of functions (charge
density/potential) supported on the cavity surface. 
\end{enumerate*}
Whereas the former
assumption is a physical one, giving a prescription for the underlying
physical laws,\cite{Jackson:1998uq} the latter is a convenient mathematical
formulation, which reduces the computational cost
transforming a three-dimensional problem in the
whole space to a two-dimensional one on the boundary of the
molecular cavity. Despite the convenience, a sharp boundary between
neighboring molecules assumes that no electronic density is
present beyond the cavity surface. This is not physically sound, because electronic densities of solute and solvent in reality overlap. Initially, this issue has been dealt with by simple renormalization procedures\cite{Tomasi:1994wt}; more elaborate corrections have later been proposed\cite{Klamt:1998ga,Chipman:1998fg,Chipman:2000eh} and for the \ac{IEF} formulation of the \ac{PCM} it can be shown that a first-order correction is already included in the model.\cite{Cances:2001eh} A full account of this issue is however not practical in terms of a surface model, and the ever
increasing basis sets employed in routine calculations, including very diffuse functions, aggravate the problem further by allowing more and more of the electron density to ``escape'' the cavity.

Neglecting electronic charge overlap between solute and solvent does not only impact the electrostatic energy: excitation energies depend on the charge distribution in the excited states, which is invariably more diffuse than in the ground state, and other interaction terms, such as the repulsion energy, depend explicitly on the overlap between solute and solvent densities.\cite{Amovilli:1997um}  

The parametric description of the cavity surface also presents challenges, not only from a formal point of view to define the correct cavity boundary,\cite{Tomasi:1994wt,Tomasi:2005ipa} but also from a technical standpoint, especially for larger
molecules. The development of stable cavity generators is still an active area of research.\cite{Silla:1990ix,Silla:1991be,PascualAhuir:1990ic,Pomelli:1998fra,Pomelli:1999bf,Connolly:1983jw,Connolly:1993fd,Foresman:1996gc,Quan2016-gm,Quan2017-zw}

In recent years, several real-space methods for quantum chemistry have
been developed,\cite{Losilla:2010ih,Genovese:2011gr,Andrade:2015cd,fhiaims,Harrison:2016ez,mrchem} and with these,
the treatment of solvation as a three-dimensional
problem has become a feasible alternative. The advantage is a seamless
integration with the quantum mechanical implementation: the
electrostatic potential is no longer computed in vacuum but in the
generalized dielectric medium with a position-dependent
permittivity. Several real-space codes have so far adopted this
strategy.\cite{FossoTande:2013ka,FossoTande:2013wv,Fisicaro:2017fs,Fisicaro:2016kl,Andreussi:2019gf,Womack:2018iu} 
Another advantage of this approach is an increased
flexibility: no constraints are placed on the form of the permittivity
function, and complex environments consisting of surfaces, droplets,
membranes, can be treated without the need of ad-hoc implementations,
which are often limited to a handful of special cases.\cite{Frediani:2000vf,Corni:2008wu,DiRemigio:2016fo}

% This is not real for us now The main disadvantage of a
%three-dimensional approach is the need to obtain the reaction
%potential through an iterative procedure, which is generally nested
%in the \ac{SCF} cycle, making the whole approach computationally
%expensive. To avoid the additional iterative procedure, we have here
%employed the variational approach developed by
%\citeauthor{Lipparini:2010}: the reaction potential is variationally
%optimized alongside the orbitals during the SCF procedure, without
%the need for a nested structure. The converged result is therefore a
%set of optimized orbitals, and the corresponding reaction potential.

In this contribution, we will present our implementation, which
makes use of a \ac{MW} framework\cite{Alpert:1993tf,Alpert2002-sr,Beylkin:2008im,Frediani:2013fb} to solve both the \ac{KS} equations of
\ac{DFT}\cite{Harrison:2004vua,Harrison:2004gd} and the \ac{GPE}\cite{FossoTande:2013ka} for the solvent reaction potential.  We will
also show a set of benchmark calculations to showcase the implementation's
theoretical correctness, parametrization, and flexibility.
\acp{MW} constitute a basis which can give accurate results up to a user-defined
precision, thanks to an automatic adaptive refinement.\cite{Frediani:2013fb} 
Our implementation is included in the open-source \ac{MW} computational chemistry software package MRChem.\cite{mrchem}
The combination of \ac{MW}-based \ac{KS}-\ac{DFT} and \ac{GPE} solver provides a methodology for the assessment of solvent effects with controlled precision.

%In Section~\ref{sec:theory} we briefly introduce the \ac{MW} framework, the formulation of the \ac{GPE}, and discuss our solution strategy for the \ac{GPE} in a real-space setting. Particular attention is devoted to the parametrization of the solute-solvent boundary and of the position-dependent permittivity.
%In Section~\ref{sec:implementation} we describe our implementation and in Section~\ref{sec:results} we present its validation with respect to sharp boundary methods.

\section{Theory}\label{sec:theory}

In the theoretical framework adopted in this work, molecules are described through quantum mechanics, whereas the solvent is modeled as a classical entity, described by macroscopic properties. The two subsystems are connected by the solute-solvent interaction which describes the mutual polarization of the two subsystems\cite{Tomasi:1994wt,Tomasi:2005ipa}. Such an interaction is described by classical electrostatics. In almost all implementations, the quantum and the classical problem are solved with very different methods: the most widely used approach makes use of \ac{BEM}\cite{Sauter2011-an} techniques to solve the electrostatic problem (environment) and \ac{GTO} bases\cite{Jensen:2013wra,Jensen:2013cr} to describe the quantum problem. The use of Multiwavelets offers a unique opportunity to treat both problems with the same tools and methods. We will here recap the basic concepts of \ac{MRA} and how it is employed to solve the electrostatic and the quantum problem.

\subsection{Multiresolution analysis and Multiwavelets}

\ac{MRA} is a mathematical framework which considers a space spanned by a basis of functions with self-similarity and regularity properties\cite{Keinert:2003uc}. In practice, all basis functions are constructed by simple translation and dilation of a small set of starting functions $\scaling(x)$:

\begin{equation}
    \scaling^{n}_{l}(x) = 2^{n/2}\scaling(2^nx - l)
\end{equation}

\begin{figure}
  \centering
  \begin{subfigure}[b]{0.35\textwidth}
      \centering
      \includegraphics[height=40mm]{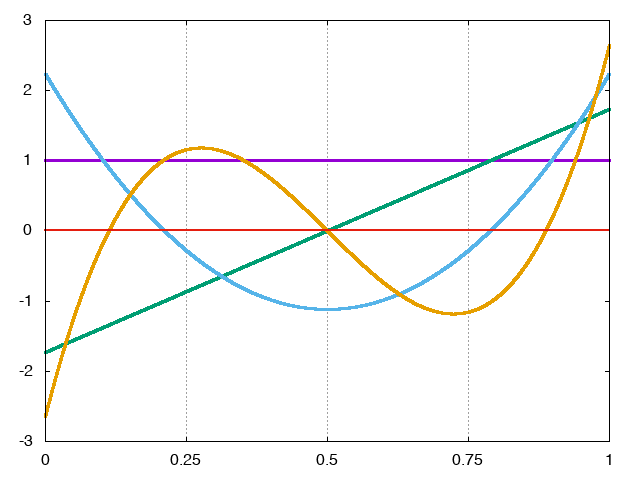}
      \caption{Scaling functions.}
      \label{fig:scaling-funs}
  \end{subfigure}
  \hfill
  \begin{subfigure}[b]{0.25\textwidth}
      \centering
      \includegraphics[height=40mm]{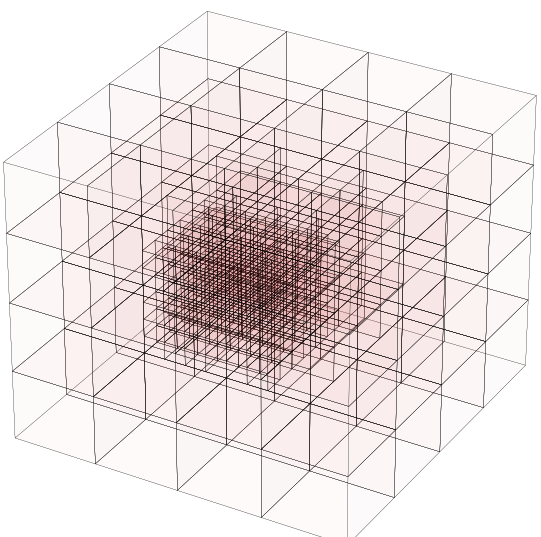}
      \caption{Adaptive grid.}
      \label{fig:adaptive-grid}
  \end{subfigure}
  \hfill
  \begin{subfigure}[b]{0.35\textwidth}
      \centering
      \includegraphics[height=40mm]{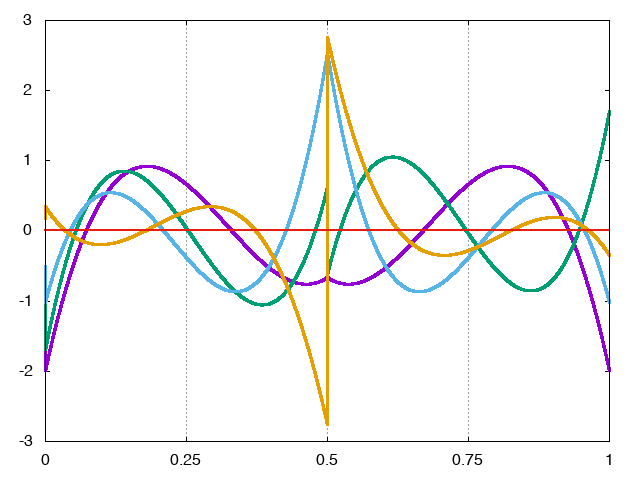}
      \caption{Wavelet functions.}
      \label{fig:wavelet-funs}
  \end{subfigure}
  \caption{Left panel: scaling functions of order $k=3$ defined in the interval $[0,1]$ are simple polynomials. 
  Right panel: the corresponding wavelet functions are piece-wise polynomials with four vanishing moments (orthogonal to polynomials up to the cubic one). 
  Central panel: adaptive grids are constructed on demand to minimize storage and meet precision requirements.}
  \label{fig:wavelets}. 
\end{figure}

The core idea of \ac{MRA} is that the space spanned by the basis functions at a given \emph{scale} $n$ is a subspace of those at scale $n+1$. Such a \emph{ladder of spaces} can be extended indefinitely and its limit is by construction dense in $L^2$. Successive refinements thus provide a systematic strategy to reach completeness, with a handful of predefined functions. This is in stark contrast with traditional \ac{GTO} methods, where extending a basis requires a complete reparametrization of the basis set, atom by atom. The \emph{wavelet} functions are obtained by taking the difference between two consecutive scaling spaces, and they convey information about the error incurred  at each scale $n$ due to neglecting the refinement at scale $n+1$, see Figure~\ref{fig:wavelets} for a 1-dimensional illustration.

As long as the fundamental properties of \emph{self-similarity} and \emph{completeness} are preserved, the choice of a specific basis set can be guided by numerical considerations to obtain compact representation of functions and efficient application of operators.

Alpert's Multiwavelets\cite{Alpert:1993tf} constitute a practical realization of \ac{MRA} by considering a set of polynomial functions (\emph{e.g.} Legendre or Interpolating polynomials) defined on an interval. The main advantages of Multiwavelets are the simplicity of the original basis (a polynomial set) and the disjoint support (basis functions are zero outside their support node)\cite{Alpert2002-sr}. The latter enables adaptive refinement of functions to minimize the storage needs and the computational overhead.
The extension to three-dimensional functions is obtained by tensor-product methods, and operators are efficiently applied in a separated form\cite{Beylkin:2008im}.

Multiwavelets are an ideal framework to deal with integral operators, and this allows both the \ac{KS} equations for the quantum system\cite{Harrison:2004vua} and the Poisson equation for the solvent polarization\cite{FossoTande:2013ka} to be solved within the same formalism, once the equations are converted from the conventional differential form to the appropriate integral form. Functions are projected/computed on an adaptive grid to guarantee the requested precision.
All operations (operator applications, algebraic manipulations) are defined within the requested precision, in such a way that the developer can easily implement new algorithms with little effort\cite{VAMPyR_paper} and the end-user only needs to specify the requested precision\cite{mrchem,Jensen:2017bv,Jensen:2016jy,Brakestad:2020gv}.

For details about how to solve the \ac{KS} equations within a \ac{MW} framework, we refer to the literature\cite{Harrison:2004vua,Frediani:2013fb,Harrison:2004gd}. Concerning the \ac{GPE} we will expose the derivation and the implementation details in the remainder of this section.

%%\begin{equation}\label{eq:scalingsubspaces}
%  V^k_0 \subset V^k_1\subset V^k_2\subset \cdots \subset V^k_n \subset \cdots.
%\end{equation}
%where $k$ is the polynomial order, $n$ is the refinement level and $V^k_0$ is spanned by a basis of
%piece-wise polynomials $\{\phi_0, \ldots, \phi_{k-1}\}$. The rest of the
%subspaces are spanned by dilation and translations of the basis of $V^k_0$. We
%define the orthogonal complement of $V^k_n$ as $W^k_n$ and
%\begin{equation}\label{eq:waveletsubspace}
%  V^k_n \oplus W^k_n = V^k_{n+1}.
%\end{equation}
%Combining statements \ref{eq:scalingsubspaces} and \ref{eq:waveletsubspace} we
%arrive at
%\begin{equation}\label{def:nestedsubspaces}
%  V^k_0 \oplus W^k_1\oplus W^k_2\oplus \cdots \oplus W^k_{n-1} = V^k_n.
%\end{equation}
%The above definition gives us a way to work with these spaces. Instead of projecting with the scaling %basis at each level, we use the orthogonal complements, the multi-wavelets, and project into the %multiwavelet spaces. This way we can check for convergence by comparing the projection coefficients with %an arbitrary user-defined threshold.

\subsection{Electrostatics of continuous media}\label{sec:macro-maxwell}
Any material is a bound aggregate of nuclei and electrons: at
microscopic level these charged particles obey the microscopic Maxwell
equations.
We are however interested in the \emph{macroscopic} behaviour of the material in
the presence of external sources of charge $\rho(\vect{r})$ and current
$\vect{j}(\vect{r})$. Following Jackson,\cite{Jackson:1998uq} we
can perform a spatial average to arrive at the \emph{macroscopic} Maxwell equations:
\begin{equation}
  \label{eq:macro-maxwell}
  \begin{cases}
   &\nabla \cdot \vect{D} = 4\pi \rho \\ 
   &\nabla \times \vect{H} - \frac{1}{c}\pderiv{\vect{D}}{t} = \frac{4\pi}{c}\vect{j} \\
   &\nabla \times \vect{E} + \frac{1}{c}\pderiv{\vect{B}}{t} = 0 \\
   &\nabla \cdot \vect{B} = 0.
  \end{cases}
\end{equation}
These equations are expressed in terms of the usual electric and magnetic
fields, $\vect{E}$ and $\vect{B}$, and additionally the \emph{displacement}
$\vect{D}$ and \emph{magnetization} $\vect{H}$ fields appear as a result of the
spatial averaging.
In the quasistatic limit, the electric field has zero curl and can thus be
written in terms of a scalar potential function: $\vect{E} = -\nabla V$,
where $V$ is the electrostatic potential. To relate the external sources to
the potential it is first necessary to relate the fields $\vect{E}$ and
$\vect{D}$ with a \emph{constitutive
  relation},\cite{Jackson:1998uq,Norman2018-tn} which is, in general, a
nonlinear and space-time nonlocal relationship between the fields. For 
linear and local continuous media the constitutive relation is:
\begin{equation}\label{eq:linear-media}
  \vect{D} = \vect{\varepsilon}(\vect{r})\vect{E},
\end{equation}
where the permittivity $\vect{\varepsilon}(\vect{r})$ is a position-dependent,
rank-3 symmetric tensor. Upon inserting the constitutive relation into the first of
Maxwell's equations we obtain the \ac{GPE}:
\begin{equation}\label{eq:tensor-gpe} 
  \nabla \cdot [\vect{\varepsilon}(\vect{r}) \nabla V] = -4\pi\rho.
\end{equation}
In the following we will further specialize to the isotropic case
$\vect{\varepsilon}(\vect{r}) = \varepsilon(\vect{r})\mat{I}$, with $\mat{I}$
the rank-3 identity:
\begin{equation}\label{eq:gpe} 
  \nabla \cdot [\varepsilon(\vect{r}) \nabla V] = -4\pi\rho.
\end{equation}
We remark that the permittivity is still position-dependent, in contrast to the usual \ac{PCM} treatment.
The solution to Eq. \eqref{eq:gpe} can be partitioned as:
\begin{equation}\label{eq:newton_plus_reaction}
  \pot 
= \newton + \potsol
= \int_{\mathbb{R}^{3}} \frac{\rho(\vect{r}^{\prime})}{|\vect{r} - \vect{r}^{\prime}|}\diff\vect{r}^{\prime} + \potsol
\end{equation}
where $\newton$ is the electrostatic potential in vacuum and $\potsol$ is the \emph{reaction potential}. The \emph{polarization energy} is then defined as:
\begin{equation}\label{eq:polarization-energy}
   U_{\mathrm{pol}} = \frac{1}{2}\int \diff \mathbf{r} \rho(\mathbf{r}) \potsol[\rho](\mathbf{r}).
\end{equation}
We write the reaction potential as a functional of the charge density: the functional dependence is linear.\cite{Cances:2001eh}

\subsection{The quantum-classical coupling}\label{sec:qm-classical}
Our quantum mechanical treatment of the system will be based on \ac{KS}-\ac{DFT}.
For an $N$-electron system coupled with a classical polarizable continuum environment, the \ac{KS}-\ac{DFT} \emph{free} energy\cite{Tomasi:2005ipa} functional\cite{Parr1994-mm,Lin2019-ox} reads:
\begin{equation}\label{eq:ks-dft-functional}
  \begin{aligned}
  \mathcal{G}[\rho] &= 
  T_{\mathrm{s}}[\rho_{\mathrm{e}}] 
  +
  V_{\mathrm{Ne}}[\rho_{\mathrm{e}}] 
  + 
  J[\rho_{\mathrm{e}}] 
  - 
  \zeta K[\rho_{\mathrm{e}}] 
  +
  E_{\mathrm{xc}}[\rho_{\mathrm{e}}, \nabla\rho_{\mathrm{e}}] 
  + 
  U_{\mathrm{pol}}[\rho] 
  + 
  \frac{1}{2}\sum_{\alpha \neq \beta}^{N_{\mathrm{nuclei}}} \frac{Z_{\alpha}Z_{\beta}}{| \vect{R}_{\alpha} - \vect{R}_{\beta}|} \\
  &=
  \int \diff \vect{r}
\left[ -\frac{1}{2} \nabla^{{2}} \rdm \right]_{\vect{r}^{\prime} = \vect{r}}
  + \int \diff \vect{r} V_{\mathrm{Ne}}(\vect{r}) \rho_{\mathrm{e}} \\
  &+ \frac{1}{2} \int \diff \vect{r} \int\diff \vect{r}^{\prime}
  \frac{\rho_{\mathrm{e}}(\vect{r}) \rho_{\mathrm{e}}(\vect{r}^{\prime})}{|\vect{r} - \vect{r}^{\prime}|}
  - \frac{\zeta}{2} \int \diff \vect{r} \int\diff \vect{r}^{\prime}
  \frac{
  \opdm{\vect{r}}{\vect{r}^{\prime}}\opdm{\vect{r'}}{\vect{r}}
  }{|\vect{r} - \vect{r}^{\prime}|}
 + E_{\mathrm{xc}}[\rho_{\mathrm{e}}, \nabla\rho_{\mathrm{e}}] \\ 
 &+ \frac{1}{2}\int \diff \mathbf{r} \rho(\mathbf{r}) \potsol[\rho](\mathbf{r})
  + 
  \frac{1}{2}\sum_{\alpha \neq \beta}^{N_{\mathrm{nuclei}}} \frac{Z_{\alpha}Z_{\beta}}{| \vect{R}_{\alpha} - \vect{R}_{\beta}|}.
 \end{aligned}
\end{equation}
The molecular charge density is separated into electronic and nuclear components:
\begin{equation}
\rho(\vect{r}) = \rho_{\mathrm{e}}(\vect{r}) + \sum_{\alpha}^{N_{\mathrm{nuclei}}} Z_{\alpha}\delta(\vect{r} - \vect{R}_{\alpha}),
\end{equation}
$E_{\mathrm{xc}}[\rho_{\mathrm{e}}, \nabla\rho_{\mathrm{e}}]$ is a GGA exchange-correlation functional, and the nuclear-electron potential is defined as:
\begin{equation}
V_{\mathrm{Ne}}(\vect{r}) = - \sum_{\alpha = 1}^{N_{\mathrm{nuclei}}} \frac{Z_{\alpha}}{| \vect{R}_{\alpha} - \vect{r}|}.
\end{equation}
$\zeta$ is a scalar factor influencing the portion of exact exchange included in the energy.
The \ac{RDM} and electronic density function appear in the energy expression:
\begin{alignat}{2}\label{eq:densities}
 \rdm = \sum_{i=1}^{{N}} \phi_{i}(\vect{r})\phi^{*}_{i}(\vect{r}^{\prime}), 
 &\quad
 \rho_{\mathrm{e}}(\vect{r}) \equiv \opdm{\vect{r}}{\vect{r}}. 
\end{alignat}

The minimum is found by constrained optimization, to enforce idempotency and normalization of the \ac{RDM}:
\begin{equation}
  \min_{\rho_{\mathrm{e}}} \mathcal{G}[\rho] \text{ such that }
  \begin{cases}
    &\int \diff\vect{r}^{\prime\prime} \opdm{\vect{r}}{\vect{r}^{\prime\prime}}\opdm{\vect{r}^{\prime\prime}}{\vect{r}^{\prime}} = \opdm{\vect{r}}{\vect{r}^{\prime}} \\
    &\int \diff\vect{r} \rho_{\mathrm{e}} = N
  \end{cases}
\end{equation}
and leads to the variational condition:\cite{Parr1994-mm,McWeeny1960-jm}
\begin{equation}\label{eq:var-condition}
  \left[ F, \rho_{\mathrm{e}} \right] = 0,
\end{equation}
where the effective one-electron Fock operator appears:
\begin{equation}\label{eq:fock-operator}
  \begin{aligned}
  F(\vect{r}, \vect{r}^{\prime})
  =
  \frac{\delta \mathcal{G}}{\delta \rdm}
  &=
  \left[ -\frac{1}{2} \nabla^{{2}} + V_{\mathrm{Ne}}(\vect{r}) \right] \dirac{\vect{r}}{\vect{r}^{\prime}} 
  +
  \dirac{\vect{r}}{\vect{r}^{\prime}}
  \left[ \int \diff \vect{r}^{\prime}\frac{\rho_{\mathrm{e}} (\vect{r}^{\prime})}{|\vect{r} - \vect{r}^{\prime}|} \right] \\
  &-
  \frac{\zeta}{|\vect{r} - \vect{r}^{\prime}|}\opdm{\vect{r}}{\vect{r}^{\prime}}
  +
  \frac{\delta E_{\mathrm{xc}}}{\delta \rdm}
  + 
  \potsol[\dens]
  \end{aligned}
\end{equation}

\subsection{Solving the generalized Poisson equation}\label{sec:solving-classical}

The solution to the \ac{GPE} is a function supported on the entire space $\mathbb{R}^3$.
Apparent surface charge formulations of continuum solvation
models do not solve Eq. \eqref{eq:gpe} directly, but rather reformulate it as a boundary integral equation and solve it by boundary-element discretization. The apparent surface charge, supported on the closed solute-solvent boundary, is the sought-after quantity to compute the polarization energy.\cite{Cances:2001eh}
Such a procedure is generally based on two underlying assumptions:
\begin{enumerate*}[label=(\arabic*)]
\item the charge density is entirely contained inside the cavity boundary, and 
\item the permittivity is unitary inside the cavity and constant outside the cavity, with a jump condition which defines the electrostatic
potential and field across the cavity boundary. 
\end{enumerate*}
With a real-space approach both
assumptions can be relaxed and the equation can be solved directly. We recap
here the procedure outlined by Fosso-Tande and Harrison\cite{FossoTande:2013ka}.

We rewrite Eq. \eqref{eq:gpe} in terms of the Laplacian of the potential \pot:
\begin{equation}\label{eq:delta_explicit}
    \nabla^{2} \pot = -\frac{4 \pi \rho}{\permittivity} - \frac{\nabla \permittivity \cdot \nabla \pot}{\permittivity}.
\end{equation}
The second term on the right-hand side contains both the gradient of the permittivity and the gradient of the potential. When the permittivity is not constant, the equation cannot be solved in one step by inversion of the Laplacian, \emph{i.e.} by convolution of the right-hand side with the Laplacian's Green's function. An iterative strategy must be employed instead.

Let us then define the effective charge:
\begin{equation}\label{eq:rhoeff}
\rhoeff{}=\frac{\rho}{\varepsilon}
\end{equation}
and the polarization function: 
\begin{equation}\label{eq:gamma}
\polarization
  = 
  \frac{1}{4\pi} 
  \frac{\nabla \varepsilon \cdot \nabla \pot}{\varepsilon}
  =
  \frac{\nabla \log \varepsilon \cdot \nabla \pot}{4 \pi}
  ,
\end{equation}
such that Eq. \eqref{eq:delta_explicit} becomes:
\begin{equation}\label{eq:delta_implicit}
    \nabla^{2} \pot = -4 \pi \left( \rhoeff + \polarization \right)
\end{equation}
We can now formally solve Eq. \eqref{eq:delta_explicit} in terms of the Laplacian's Green's function:
\begin{equation}\label{eq:green}
    \pot(\vect{r})
    = 
    \int \diff \vect{r}^\prime \frac{\rhoeff(\vect{r}^\prime) + \polarization(\vect{r}^\prime)}{|\vect{r} - \vect{r}^\prime|} 
    =
    \frac{1}{|\vect{r} - \vect{r}^\prime|} \star \left( \rhoeff + \polarization \right)
\end{equation}
However, both the polarization energy in Eq.\eqref{eq:polarization-energy} and the solute-solvent interaction term in the Fock operator are expressed in terms of the reaction potential, rather than the total electrostatic potential.
By making use of the partition of $V$ in Eq. \eqref{eq:newton_plus_reaction} and recalling that $\nabla^{2} V_{\rho} = - 4\pi\rho$ one obtains
\begin{equation}
    \nabla^{2}\potsol = -4 \pi 
    \left[ \rho \left(\frac{1-\varepsilon}{\varepsilon}\right) + \polarization \right],
\end{equation}
which can be formally inverted using the Poisson kernel:
\begin{equation}\label{eq:potsol-iter}
  \potsol
    = 
    \frac{1}{|\vect{r} - \vect{r}^\prime|} \star 
    \left[ 
    \rho \left(\frac{1-\varepsilon}{\varepsilon}\right) + \polarization 
    \right]
\end{equation}
We stress that \polarization~is a function of $V=V_{\rho}+\potsol$ and Eq.~\eqref{eq:potsol-iter} must therefore be solved iteratively.

\section{Implementation}\label{sec:implementation}
In this section we present details about our specific choice of parametrization for the permittivity and how we compute the electrostatic potential between solute and solvent. We also show how we couple this to a standard \ac{SCF} optimization procedure. 

\subsection{The permittivity function parametrization}

We partition space into two regions: a cavity containing the solute, and the remainder.
The cavity surface is defined as the union set of a collection of interlocking spheres centered on the nuclei. Their radii are parametrized by using the corresponding van der Waals radii times a factor. This factor is often set to either 1.1 or 1.2\cite{Tomasi:2005ipa}, but it might vary \emph{e.g.} depending on the charge of the solute. For standard continuum models the cavity boundary is the support of the electrostatic problem for the solute-solvent interaction. In the current model it serves as a support to define the parametrization of the position-dependent $\varepsilon(\vect{r})$. In Section~\ref{sec:results} the appropriate parametrization of the cavity for the present model will be discussed. 

Following \citeauthor{FossoTande:2013ka}, we write the permittivity as a function of the molecular cavity function:\cite{FossoTande:2013ka}
\begin{equation}\label{eq:expperm}
    \permittivity = 
    \varepsilon_{\mathrm{in}}
    \exp
    \left[
      \left( \log\frac{\varepsilon_{\mathrm{out}}}{\varepsilon_{\mathrm{in}}}\right) 
      \bigg( 1 - C(\vect{r}) \bigg)
    \right].
\end{equation}
The exponential parametrization proves convenient in light of the definition of $\gamma$ in Eq. \eqref{eq:gamma} which lets us define its gradient using the cavity function, $C(\vect{r})$, only.

\begin{figure}[!ht]
  \centering
  \includegraphics[width=0.9\textwidth]{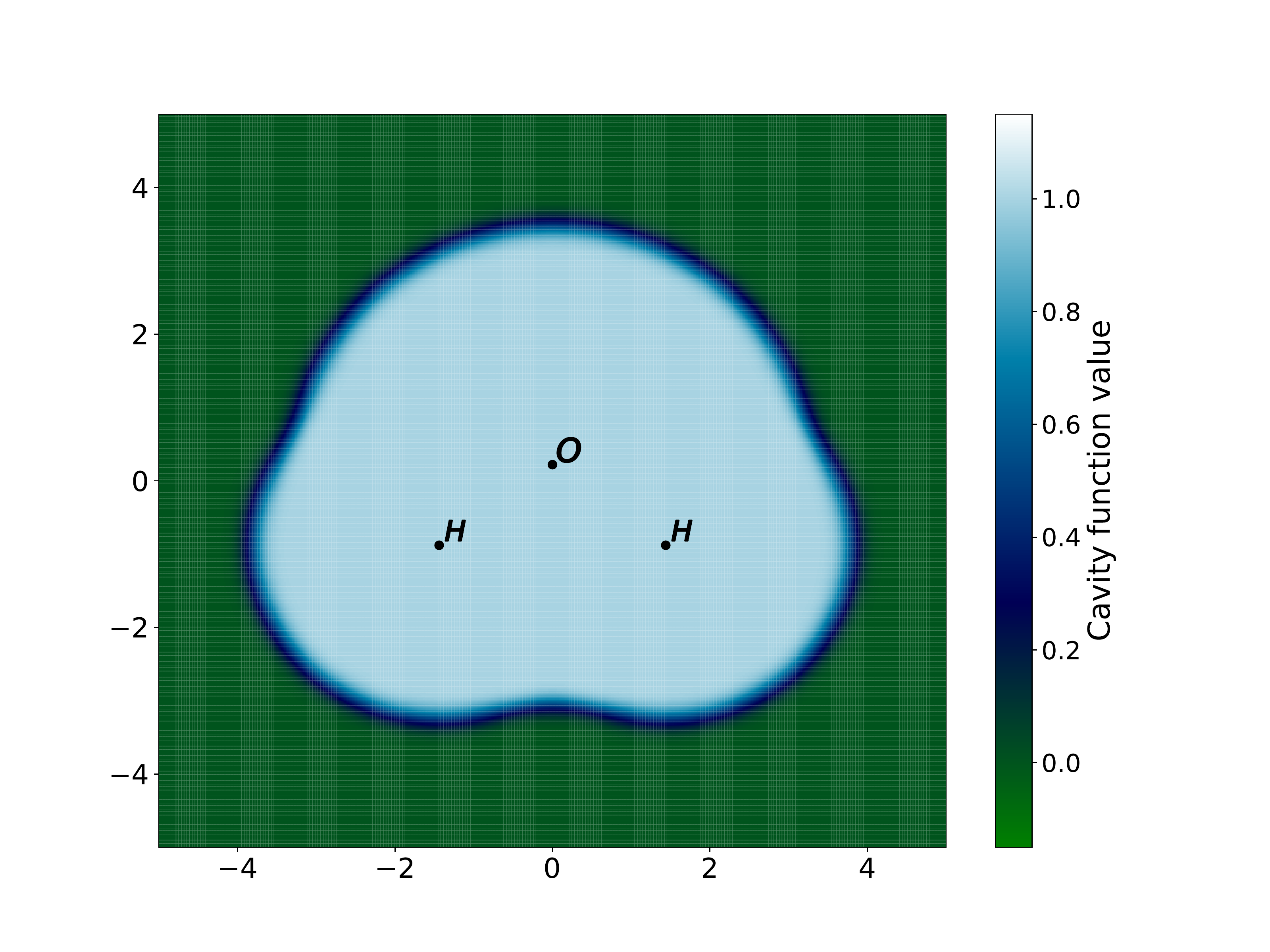}
  %\includesvg[inkscapelatex=false,width=.6\textwidth]{cavity.svg}
  \caption{Cross-section in the $xy$ plane of the cavity function $C(\vect{r})$ for the water molecule. Atom positions are indicated by their symbol. Coordinates are in atomic units. We can observe the smooth boundary of the cavity function.}
  \label{fig:cavity-h2o}
\end{figure}

The molecular cavity function is constructed as follows. For each sphere $\alpha$ centered at $\vect{r}_{\alpha}$ with radius $R_{\alpha}$, we can measure the signed normal distance of any point in space as:
\begin{equation}
    s_{\alpha}(\vect{r}) = \left|\vect{r} - \vect{r}_{\alpha}\right| - R_{\alpha}.
\end{equation}
Given $s_{\alpha}(\vect{r})$, we define a smoothed boundary of the sphere as:
\begin{equation}\label{eq:C-alpha}
    C_{\alpha}(\vect{r}) = \frac{1}{2}\left[ 1 + \erf\left(-\frac{s_{\alpha}(\vect{r})}{\sigma}\right) \right],
\end{equation}
where $\sigma$ is a user-defined smoothing parameter: $C_{\alpha}$ approaches the Heaviside step function as $\sigma \rightarrow 0$.
The molecular cavity function is then a product of all $N$ spheres:
\begin{equation}
    C(\vect{r}) = 1 - \prod^{N_{\mathrm{sph}}}_{\alpha=1} \left(1 - C_{\alpha}(\vect{r})\right),
\end{equation}
see Figure~\ref{fig:cavity-h2o} for an example.

The log-derivative of the permittivity in Eq. \eqref{eq:gamma} is then:
\begin{equation}
  \nabla\log\permittivity 
  = 
  \left(\log\frac{\epsilon_{\mathrm{in}}}{\epsilon_{\mathrm{out}}}\right)\nabla C(\vect{r}),
\end{equation}
requiring evaluation of the gradient of the cavity function. 
For interlocking-spheres cavities, a closed-form analytical expression is available, see Appendix \ref{app:eps-and-C-derivatives}, and is implemented in our code.
Note however that, in a real-space, multiwavelet framework, we can compute this gradient by direct application of the derivative operator\cite{Anderson2019-bx}, which allows to use more complex or even numerical definitions of the boundary, \emph{e.g.} as isodensity surfaces.

\subsection{The \acl{SCRF}}
The \ac{SCRF} is the iterative procedure to solve the \ac{GPE} for any given molecular density. 
At convergence, the iterations produce the reaction potential $\potsol$, which can be directly employed in the solution of the \ac{KS}-\ac{DFT} equations.

Algorithm~\ref{alg:SCRF-iter} shows the iterative procedure implemented to solve the \ac{GPE} within the \ac{SCF} iterations. The input parameters at iteration $n$ are the charge density $\rho^{[n]}$, the permittivity $\varepsilon(\vect{r})$, a guess for the reaction potential $\potsol^{[n,0]}$, and a threshold parameter $\delta$. Before iterating, the effective density $\rhoeff^{[n]}$ and the potential $\newton^{[n]}$ are computed. At each microiteration $i$, the reaction potential $\potsol^{[n,i]}$ is computed in four steps as outlined in lines 5-8 of Algorithm 1, and convergence in the norm of the reaction potential is checked against the threshold $\delta$.
At the first \ac{SCF} iteration the starting guess for the reaction potential is set to zero ($\potsol^{[0,0]} = 0$). At all subsequent iterations, the starting guess is set to the converged reaction potential from the previous iteration: $\potsol^{[n,0]} = \potsol^{[n-1]}$.

\begin{algorithm}
  \caption{Self-consistent optimization of the reaction field. $n$ is the \ac{SCF} iteration index.}\label{alg:SCRF-iter}
  \begin{algorithmic}[1]
    \Procedure{SCRF microiteration}{$\rho^{[n]}$, $\permittivity$, $\potsol^{[n,0]}$, $\delta$}
    \State $\rhoeff^{[n]} = \frac{\rho^{[n]}}{\varepsilon}$
    \State $\newton^{[n]} = \frac{1}{|\vect{r} - \vect{r}^\prime|} \star \rho^{[n]}$
    \While{$i < N_{\mathrm{micro}}$}
    \State $\pot = \potsol^{[n,i]} + \newton^{[n]}$
    \State $\polarization = \frac{\nabla \log \permittivity \cdot \nabla \pot^{[n,i]}}{4\pi}$
    \State $\potsol = \frac{1}{|\vect{r} - \vect{r}^\prime|} \star
    \left[ 
    \rhoeff^{[n]} - \rho^{[n]} + \polarization 
    \right]$
    \State $\potsol^{[n,i+1]} = \text{KAIN}(\potsol, \potsol^{[n,i]}, \ldots \potsol^{[n,i-k]})$
    \If {$||\potsol^{[n, i+1]}-\potsol^{[n,i]}|| < \delta$}
      \State \Return $\potsol^{[n]} := \potsol^{[n, i+1]}$ 
    \EndIf 
    \EndWhile
    \EndProcedure
  \end{algorithmic}
\end{algorithm}

%The guess of the molecular density $\rho^{[0]}$ is usually of poor quality at the beginning of the \ac{SCF} procedure, hence we first perform one iteration of the solver for the integral \ac{KS} equations \emph{in vacuo}, omitting the reaction potential from the Fock operator.
%Alternatively, one could first converge a gas phase calculation and then use the optimized orbitals as initial guess for the \ac{SCF} in solution. Our experience shows that this is not needed.

A straightforward implementation of the microiterations suffers from slow convergence of the reaction potential, thus adding a significant prefactor to each \ac{SCF} iteration. We use the \ac{KAIN} method\cite{Harrison:2004gd}, which is a convergence acceleration technique, similar to Pulay's DIIS\cite{Pulay:1980jn} and Anderson's mixing\cite{Anderson:1965bv}.
At each microiteration $i$, the updated reaction potential $\potsol^{[i+1]}$ is constructed as a linear combination, with constraints, of $N$ previous iterates. The KAIN history length $N$ impacts both convergence and memory: $N=5$ is generally a good compromise between fast convergence (fewer iterations) and acceptable memory footprint.

 The \ac{KAIN} acceleration is combined with an adaptive threshold to improve the convergence rate of the microiterations: instead of converging the reaction potential to the same predefined threshold $\epsilon$ used for the  orbitals, we make use of a threshold, $\delta$, chosen to be the norm of the orbital update in the parent \ac{SCF} macroiteration. $\delta$ is thus updated during the \ac{SCF} procedure.
There are two parameters that affect the convergence pattern of the reaction potential, $\potsol$:
\begin{enumerate}
\item The guess for $\potsol$ at the start of the microiterations:
    \begin{enumerate*}[label={(\textbf{\Alph*})}]
\item $\potsol^{[n, 0]} = 0$, or
\item $\potsol^{[n, 0]} = \potsol^{[n-1]}$ (and zero for the first microiteration embedded in the first macroiteration).
    \end{enumerate*}
\item The convergence threshold for the microiterations:
    \begin{enumerate*}[resume, label={(\textbf{\Alph*})}]
\item fixed threshold $\delta$, or
\item dynamic threshold $\delta^{[n]} = |\Delta \rho^{[n]}|$.
    \end{enumerate*}
\end{enumerate}
These lead to four possible convergence regimes: \textbf{AC}, \textbf{BC}, \textbf{AD}, \textbf{BD}; the latter being our default.

\begin{figure}[!ht]
  \centering
  \includegraphics[width=\textwidth]{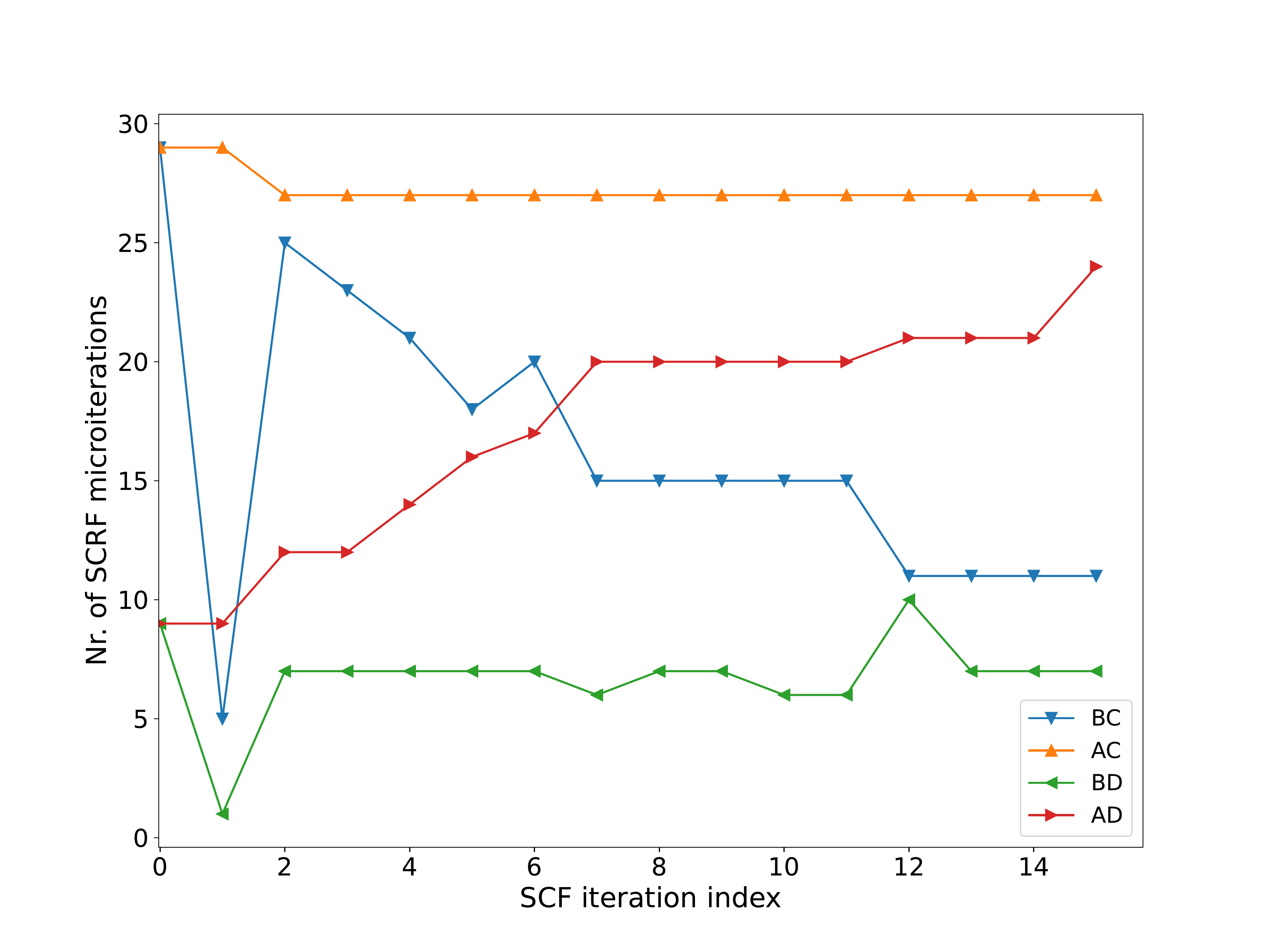}
  \caption{Convergence regimes for the \ac{SCRF} algorithm. \ac{MW} calculations with global precision $10^{-5}$ for acetamide (\ce{C2H5NO}, identifier \texttt{0233ethb} from the \acl{MSDD}). Four possible convergence scenarios are presented: static (\textbf{A}) or dynamic (\textbf{B}) precision threshold for the microiterations; zero initial guess (\textbf{C})  or guess from previous macroiteration (\textbf{D}). A dynamic threshold (green and red curves) reduces the number of microiterations at the beginning of the SCF procedure. A starting guess from the previous SCF macroiteration (green and blue curves) is effective close to convergence. Combining the two (green curve) is the optimal strategy. The dip observed for the blue and green curves at macroiteration 1, is due to the fact that the macroiteration 0 is a preliminary step and the orbital are not changed progressing from macroiteration 0 to macroiteration 1, but the convergence threshold is tightened.  This results in an almost converged reaction potential as a starting guess for the microiterations nested in macroiteration 1.}
  \label{fig:convergence-data}
\end{figure}

Figure \ref{fig:convergence-data} illustrates how the number of microiterations evolves. A dynamic precision threshold \textbf{D} reduces the number of microiterations in the beginning of the SCF procedure, simply because the threshold for convergence is looser. Using the converged $\potsol$ from the previous macroiteration \textbf{B} helps close to SCF convergence, because the orbitals do not change much and the starting guess for the microiterations is also better. Combining those two choices results in the optimal convergence pattern: the convergence threshold is progressively tighter, while at the same time the starting guess for the reaction potential improves. The opposite choice (\textbf{AC} instead of \textbf{BD}) requires a large number of microiterations throughout, whereas the intermediate choices (\textbf{AD} and \textbf{BC}) result in a large number of iterations at the beginning (\textbf{BC}) or at the end (\textbf{AD}).
We underline that all four choices converge to the same result for the example in Figure \ref{fig:convergence-data}, but we can envisage cases where convergence could potentially be prevented by choices \textbf{A} and \textbf{C}.

\section{Results}\label{sec:results}

For all systems, the solvation energies have been computed with both Gaussian16\cite{g16} and MRChem. Gaussian16 features the \ac{IEFPCM}\cite{Scalmani:2010ela} with a sharp cavity boundary. MRChem features the solvation model described in the previous sections.

Two sets of calculations have been performed. The aim of the first set was to determine a good parametrization for the cavity surface in terms of the atomic radii and the cavity surface thickness. Once a satisfactory parametrization was achieved, an extensive benchmark of solvation energies was performed, by considering the \ac{MSDD} of \citet{Marenich2020Minnesota2012}

All calculations reported are KS-DFT using the PBE0 functional.\cite{Adamo:1999hv}  Gaussian16 results are obtained with the \verb|Def2-TZVP|,\cite{weigend2005a,weigend2006a,peterson2003b} basis set, except where otherwise stated. MRChem results are obtained setting the global precision parameter to $10^{-5}$. In other words, the obtained absolute energy is correct with at least five digits with respect to the \ac{CBS} limit\cite{Jensen:2017bv}. This is not to be confused with the convergence threshold of a SCF calculation performed with an atomic basis set, which will guarantee the ``exact" result within the chosen basis, but where the precision compared to the \ac{CBS} is limited by the choice of basis.

\subsection{Cavity parametrization}\label{sec:cavity-parametrization}

For the parametrization calculations, 4 molecules of different levels of polarity were chosen: water, ethanol, formaldehyde and ethyne (geometries taken from the \ac{MSDD}\cite{Marenich2020Minnesota2012}, file names \verb|0217wat|, \verb|0045eth|, \verb|0069met| and \verb|0030eth|). No geometry optimization was performed. They were chosen to give a minimal set of neutral (polar and apolar) systems, to allow for a reliable yet simple data set to identify a good choice of the parameters defining the cavity.

 In Gaussian16, the external iteration procedure \cite{Improta:2006fza,Improta:2007dna} was used to extract the reaction energy from the total energy.\footnote{We later learned that the, undocumented, keyword \texttt{PrintResultsTable} achieves the same purpose. We used this for one molecule in the benchmark set, the singly charged negative peroxide ion \ce{O2^-} (identifier: \texttt{i091}), where the external iteration procedure failed to terminate.} The spheres used for the cavities were atom-centered and used the atoms' Bondi radii\cite{Bondi:1964fa} scaled by a factor of 1.1, as is standard for Gaussian16. Three different permittivities have been employed: 2.0, 4.0 and 80.0.  

In MRChem, the cavity is also built from atom-centered spheres, with each radius $R_{i}$ parametrized as:
\begin{equation}
    R_i = \alpha_{i} R_{i}^{\text{vdW}} + \beta_{i} \sigma_{i},
\end{equation}
where $R_{i}^{\text{vdW}}$ is the Bondi radius\cite{Bondi:1964fa,Mantina:2009gn} of the $i$-th atom, $\sigma_{i}$ is the width of the cavity boundary,
and $\alpha_{i}$ and $\beta_{i}$ are adjustable parameters. 
We allowed for granular, sphere-by-sphere flexibility in our implementation of the cavity function. By default, one value is used for each parameter ($\alpha$, $\beta$, $\sigma$) for \emph{all} spheres.
The combination $\alpha=1.1$ and $\beta=0.0$ would yield matching radii between MRChem and Gaussian16.
In the following, we explored results when $\alpha$ values were 1.0, 1.1, 1.2, 1.3 and for $\beta$ values of 0.0, 0.5, 1.0, 1.5.
In all MRChem calculations the width parameter was fixed to $\sigma = 0.2\,\text{a.u.}$ 

The aim of the parametrization is to see how the cavity width $\sigma$ affects the results of our calculations, compared to a sharp-boundary method, and to choose the combination of $\alpha$ and $\beta$ coefficients that provides a good correlation between our method and a sharp boundary implementation. The goal is not to replicate results from Gaussian16 implementation: 
our method has a diffuse cavity layer, whereas the cavity of \ac{IEFPCM} is a 2-dimensional boundary. This will lead to contributions and errors that are not equivalent.

\begin{figure}[!p]
    \centering
    %\begin{subfigure}[h]{0.24\textwidth}
    %    \centering
    %    \includegraphics[width=\textwidth]{smallparam/a10_p2_corr}
    %    
    %    \label{fig:smallparam_a10_p2}
    %\end{subfigure}
    %\hfill
    \begin{subfigure}[h]{0.40\textwidth}
        \centering
        \includegraphics[width=\textwidth]{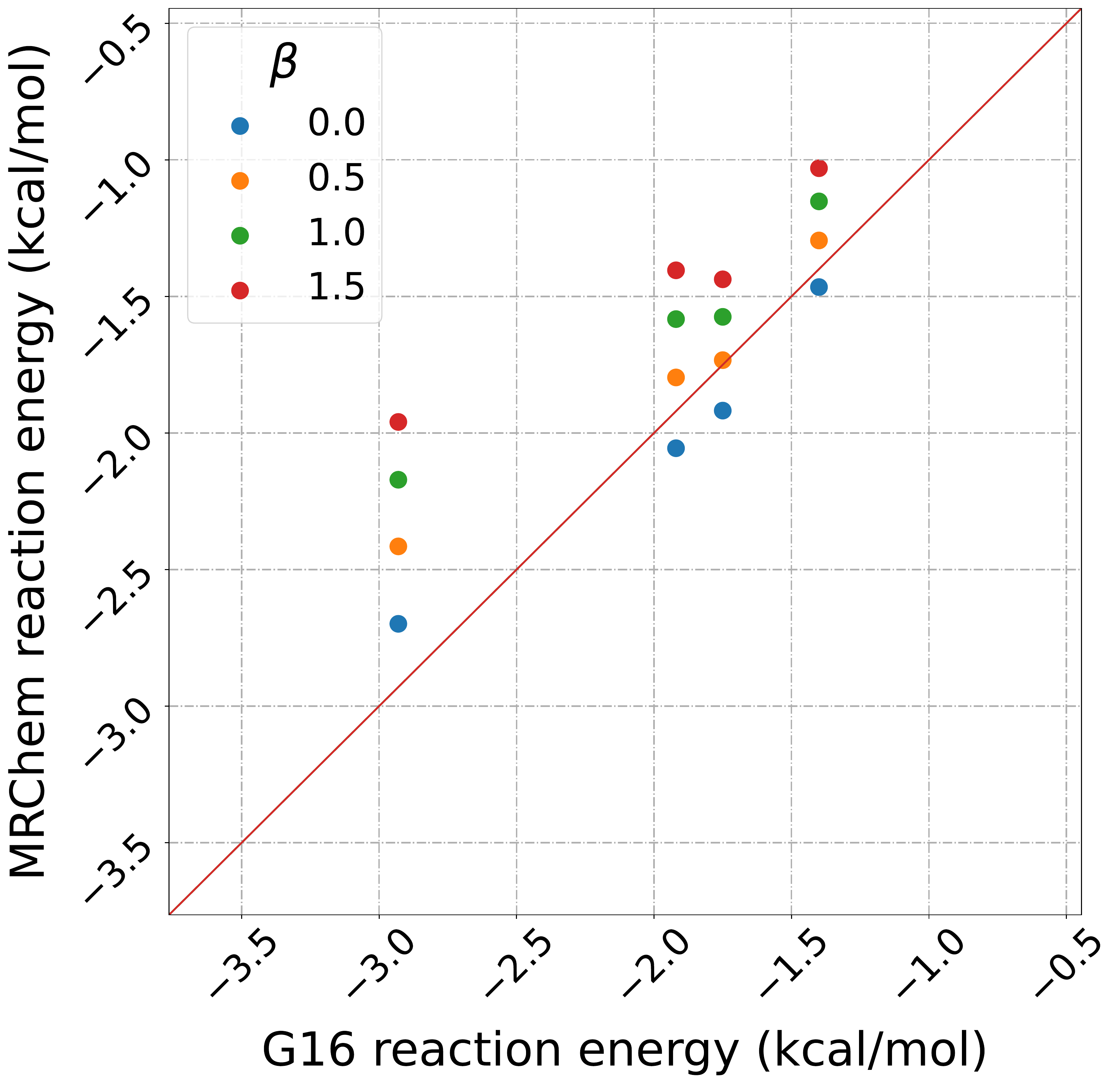}
        \caption{$\alpha = 1.1, \varepsilon=2.0$}
        \label{fig:smallparam_a11_p2}
    \end{subfigure}
    \quad\quad
    \begin{subfigure}[h]{0.40\textwidth}
        \centering
        \includegraphics[width=\textwidth]{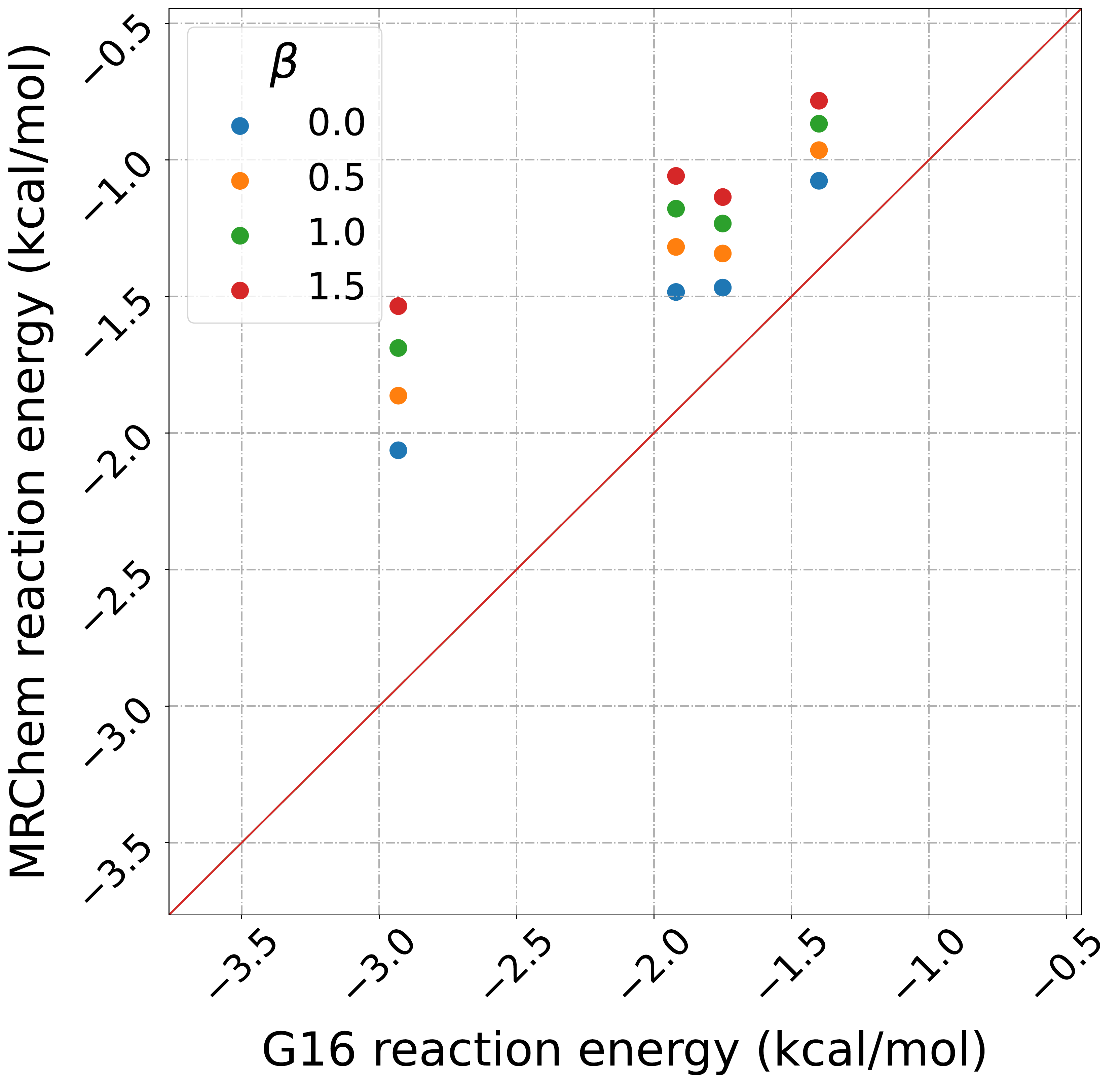}
        \caption{$\alpha = 1.2, \varepsilon=2.0$}
        \label{fig:smallparam_a12_p2}
    \end{subfigure}
    %\hfill
    %\begin{subfigure}[h]{0.24\textwidth}
    %    \centering
    %    \includegraphics[width=\textwidth]{smallparam/a13_p2_corr}
    %    
    %    \label{fig:smallparam_a13_p2}
    %\end{subfigure}
    \\
    %\begin{subfigure}[h]{0.24\textwidth}
    %    \centering
    %    \includegraphics[width=\textwidth]{smallparam/a10_p4_corr}
    %    
    %    \label{fig:smallparam_a10_p4}
    %\end{subfigure}
    %\hfill
    \begin{subfigure}[h]{0.40\textwidth}
        \centering
        \includegraphics[width=\textwidth]{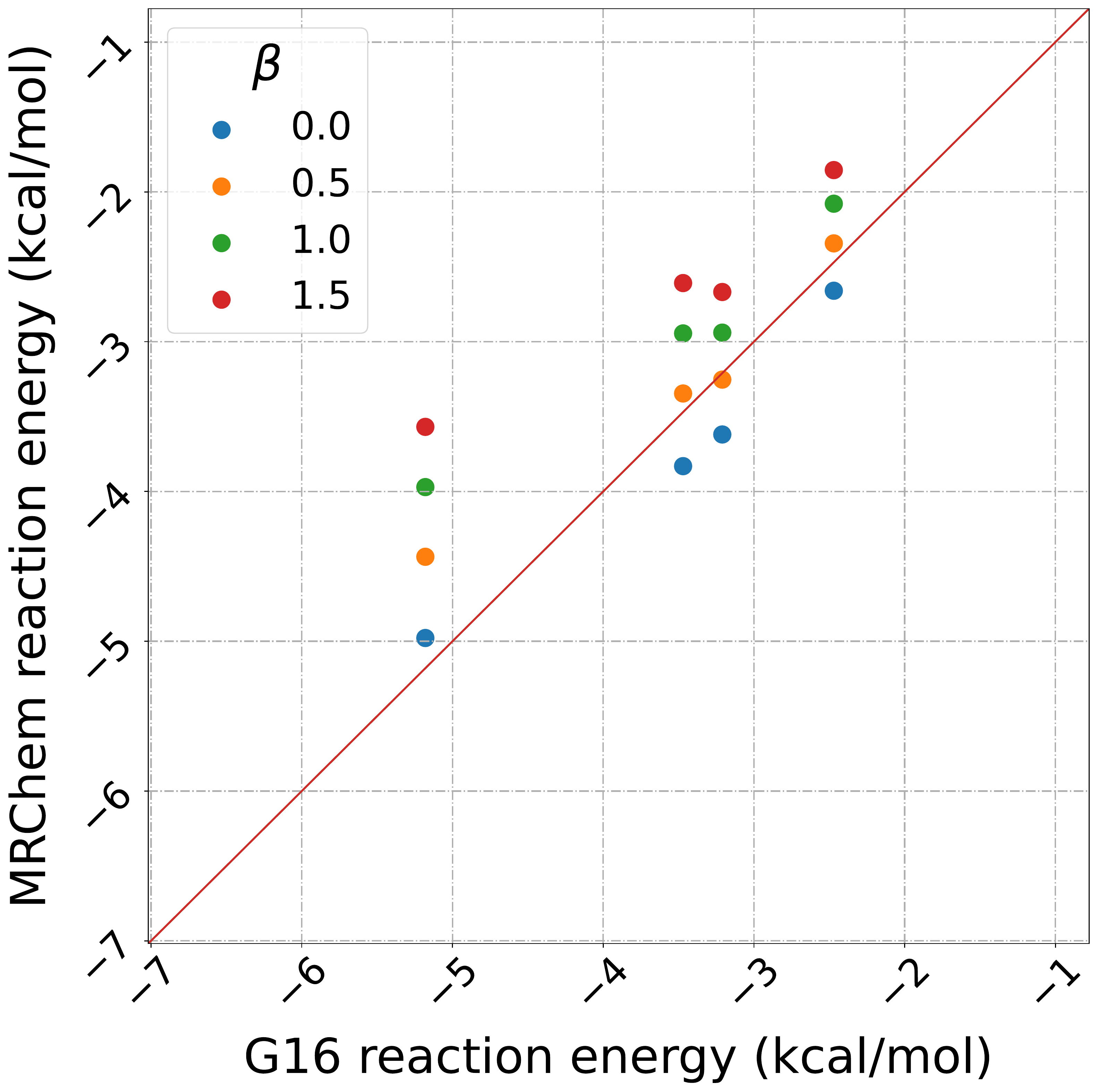}
        \caption{$\alpha = 1.1, \varepsilon=4.0$}
        \label{fig:smallparam_a11_p4}
    \end{subfigure}
    \quad\quad
    \begin{subfigure}[h]{0.40\textwidth}
        \centering
        \includegraphics[width=\textwidth]{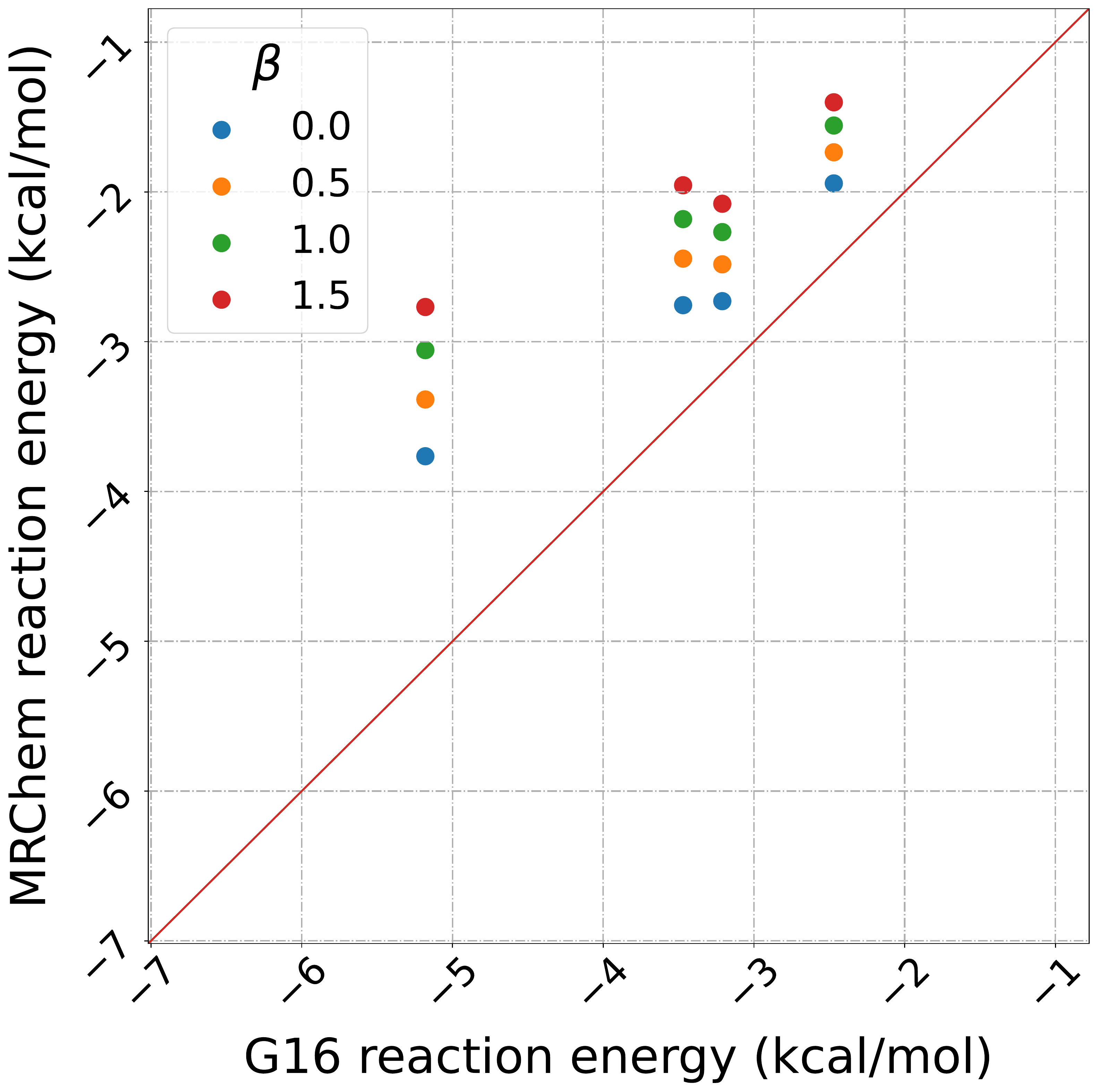}
        \caption{$\alpha = 1.2, \varepsilon=4.0$}
        \label{fig:smallparam_a12_p4}
    \end{subfigure}
    %\hfill
    %\begin{subfigure}[h]{0.24\textwidth}
    %    \centering
    %    \includegraphics[width=\textwidth]{smallparam/a13_p4_corr}
    %    
    %    \label{fig:smallparam_a13_p4}
    %\end{subfigure}
    \\
    %\begin{subfigure}[h]{0.24\textwidth}
    %    \centering
    %    \includegraphics[width=\textwidth]{smallparam/a10_p80_corr}
    %    
    %    \label{fig:smallparam_a10_p80}
    %\end{subfigure}
    %\hfill
    \begin{subfigure}[h]{0.40\textwidth}
        \centering
        \includegraphics[width=\textwidth]{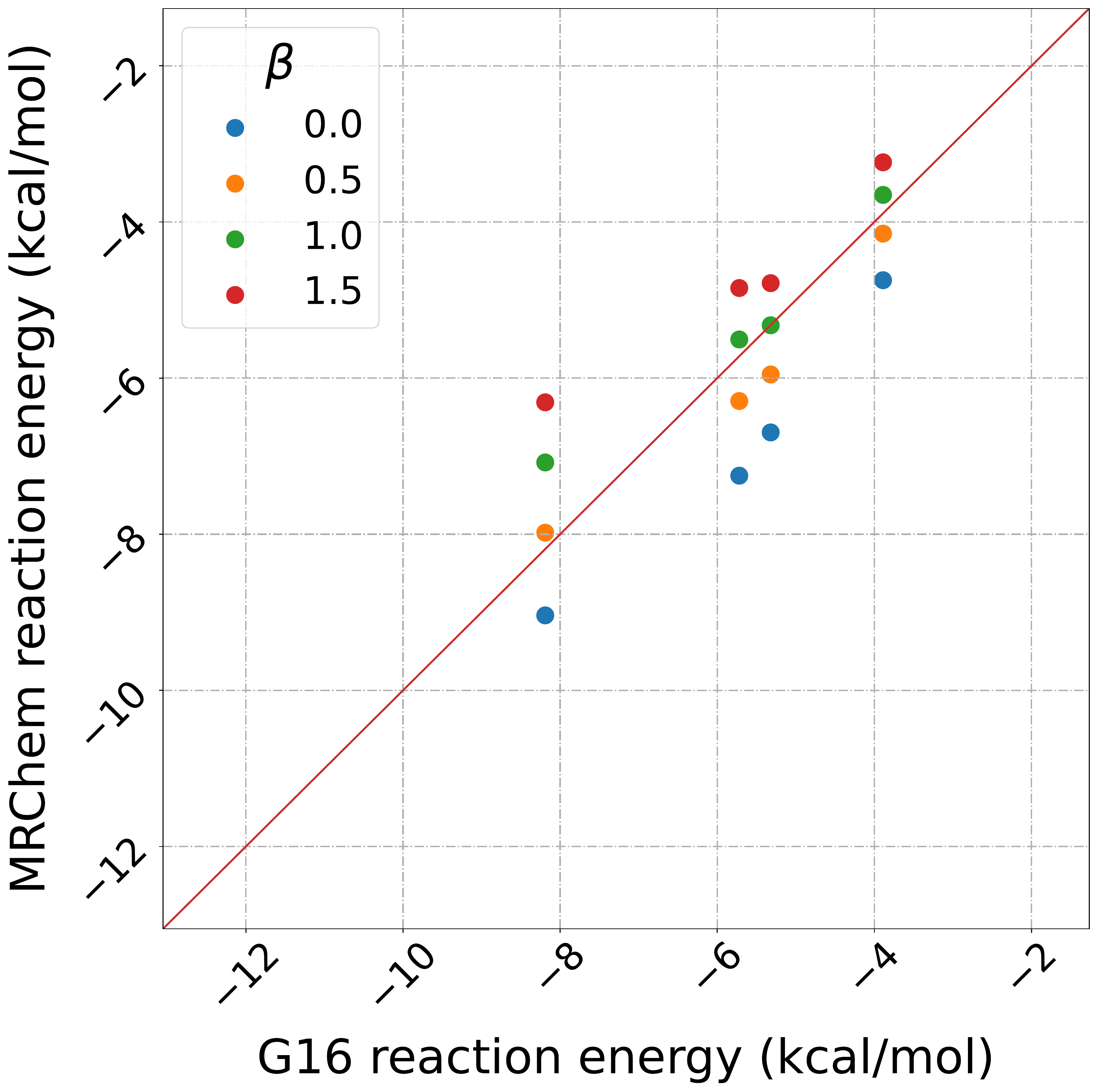}
        \caption{$\alpha = 1.1, \varepsilon=80.0$}
        \label{fig:smallparam_a11_p80}
    \end{subfigure}
    \quad\quad
    \begin{subfigure}[h]{0.40\textwidth}
        \centering
        \includegraphics[width=\textwidth]{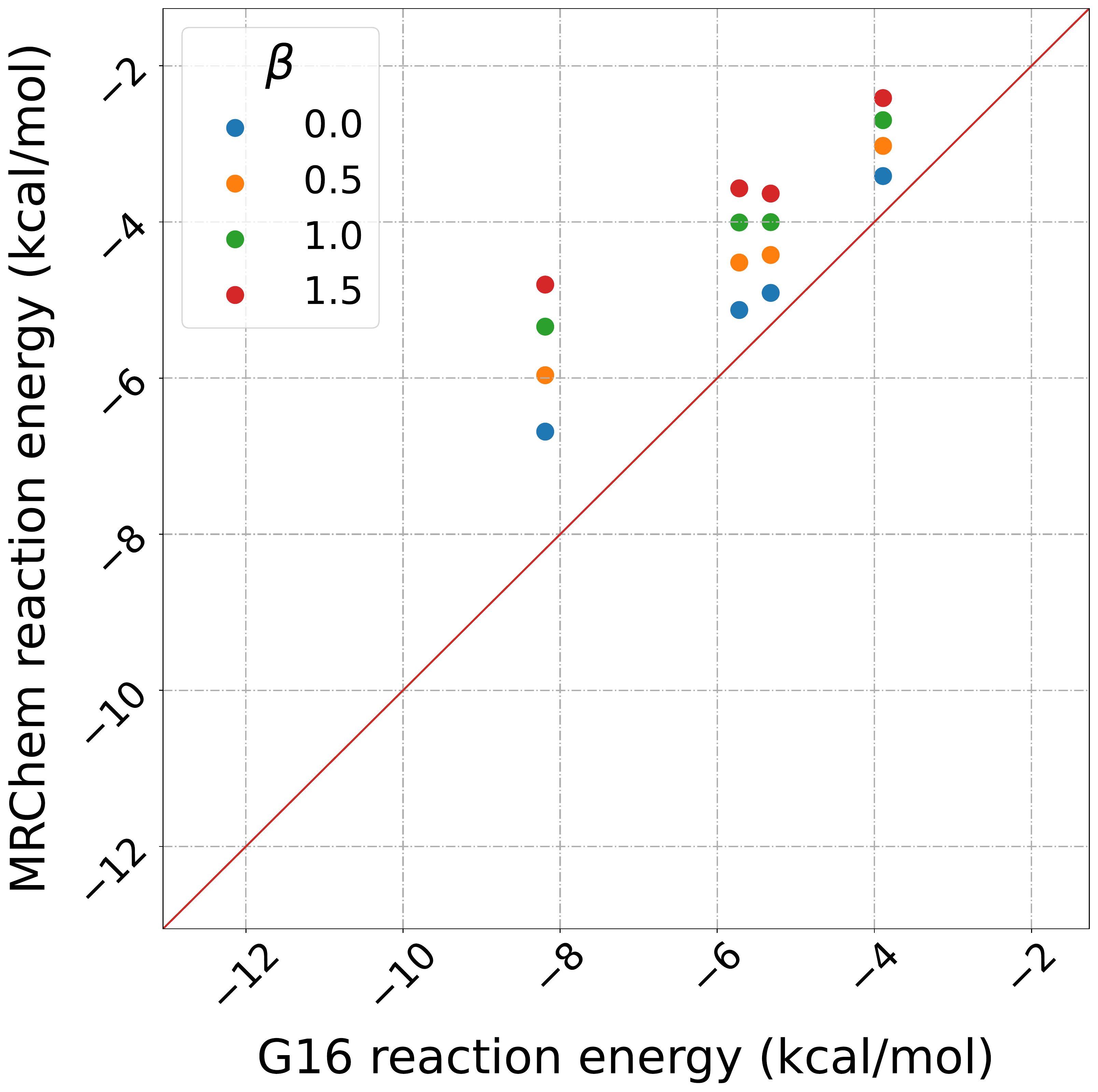}
        \caption{$\alpha = 1.2, \varepsilon=80.0$}
        \label{fig:smallparam_a12_p80}
    \end{subfigure}
    %\hfill
    %\begin{subfigure}[h]{0.24\textwidth}
    %    \centering
    %    \includegraphics[width=\textwidth]{smallparam/a13_p80_corr}
    %    
    %    \label{fig:smallparam_a13_p80}
    %\end{subfigure}
    \caption{Results for the cavity parametrization. Left column: $\alpha=1.1$. Right column: $\alpha=1.2$. On each row a different permittivity is used: from top to bottom: $\varepsilon = 2.0, 4.0, 80.0$. For each plot there are four sets of data, corresponding to $\beta=0.0, 0.5, 1.0, 1.5$. Each point on the set represents a molecule. $x$-axis: the reaction energy calculated using Gaussian16. $y$-axis: the reaction energy calculated using MRChem. Values are in Hartree. }
    \label{fig:smallparam}
\end{figure}

% show and describe results
Figure~ \ref{fig:smallparam} shows the results for the cavity parametrization for $\alpha=1.1$ and $\alpha=1.2$. Results for $\alpha=1.0$ and $\alpha=1.3$ are not shown, because they largely overestimate ($\alpha=1.0$) or underestimate ($\alpha=1.3$) solvation energies, but they are available in the data package available online on DataVerse \cite{TFSWLC_2022}.

%Running calculation with an $\alpha$ that is less than $1.1$ is possible in the current implementation, and has been done in the supplementary information, but we did not showcase it as it unphysically amplifies the effect of the continuum on the molecular density. 
%The calculations with $\alpha=1.3$ where not included in this analysis as these underestimate, for all values of the $\beta$ parameter, the reaction energy.
%This implies that values of $\alpha$ larger than 1.2 should not be used in combination with our \ac{SCRF} implementation.

We conclude that a cavity parametrization with $\alpha=1.1$ and $\beta=0.5$ provides a good correlation with sharp-boundary \ac{IEFPCM} for all reasonable values of the permittivity and default value of cavity width.  This choice of $\alpha$ and $\beta$ with $\sigma=0.2\,\mathrm{a.u.}$ is the current default in MRChem.

%In the Born model the reaction energy is inversely proportional to the radius of the cavity, this can be seen for larger radii with increasing $\alpha$ and $\beta$, where the reaction energy becomes more positive, that is, there is less interaction between the solute and solvent. 

%With all of that in mind, we chose to use the combination of parameters $\alpha=1.1$ and $\beta=0.5$ since, from our small set of calculations, is what seemed to fit best.

\subsection{Model benchmarking against the \acl{MSDD}}
The geometries from the \ac{MSDD} were used to compile a comprehensive benchmark of our model against a sharp-boundary cavity implementation. \ac{MSDD} holds solvation-related quantities, for a wide variety of solvents and solutes.\cite{Marenich2020Minnesota2012} 
From the conclusions in the previous section, all MRChem results reported in this section employ the cavity parameters $\alpha = 1.1$, $\beta=0.5$, and $\sigma =0.2\,\text{a.u.}$

\begin{figure}[!hp]
    \centering
    \begin{subfigure}[t]{0.45\textwidth}
        \centering
        \includegraphics[width=\textwidth]{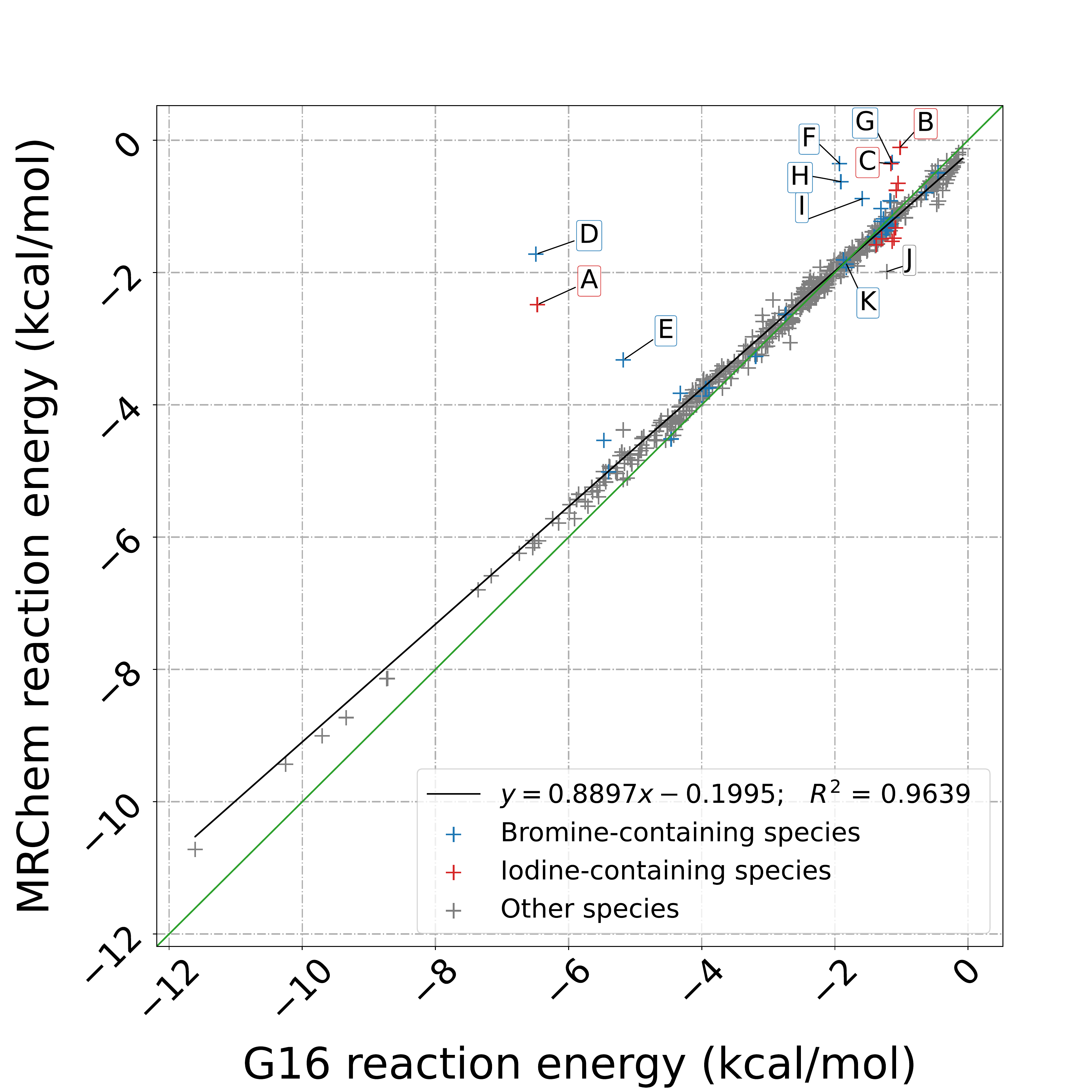}
        %\includesvg[inkscapelatex=false,width=\textwidth]{large_corr_plt_eps2_neutral.svg}
        \caption{$\varepsilon=2.0$}
        \label{fig:bigparam_a11_b5_c0_p2}
    \end{subfigure}
    \quad\quad
    \begin{subfigure}[t]{0.45\textwidth}
        \centering
        \includegraphics[width=\textwidth]{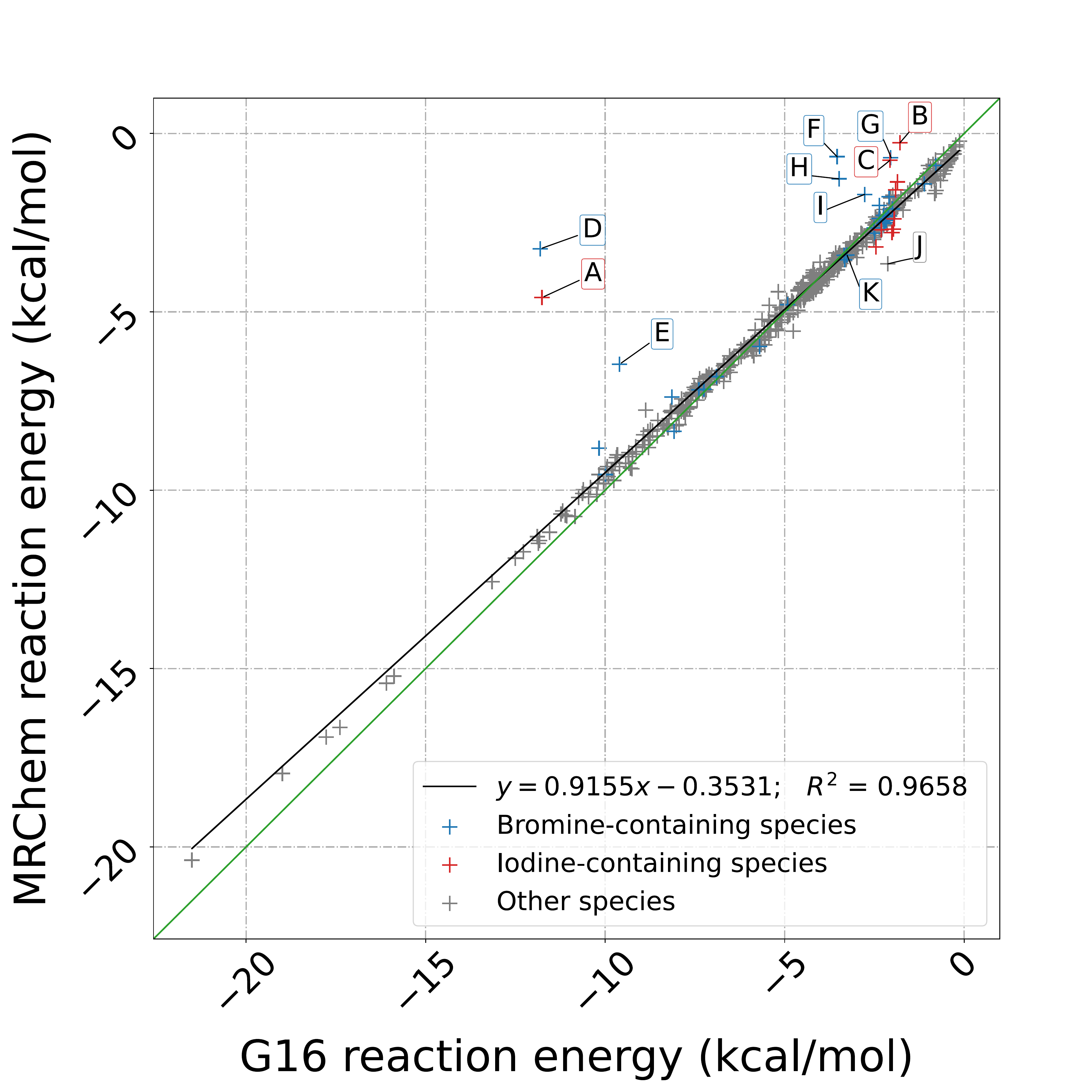}
        %\includesvg[inkscapelatex=false,width=\textwidth]{large_corr_plt_eps4_neutral.svg}
        \caption{$\varepsilon=4.0$}
        \label{fig:bigparam_a11_b5_c0_p4}
    \end{subfigure}
    \\
    \begin{subfigure}[t]{0.45\textwidth}
        \centering
        \includegraphics[width=\textwidth]{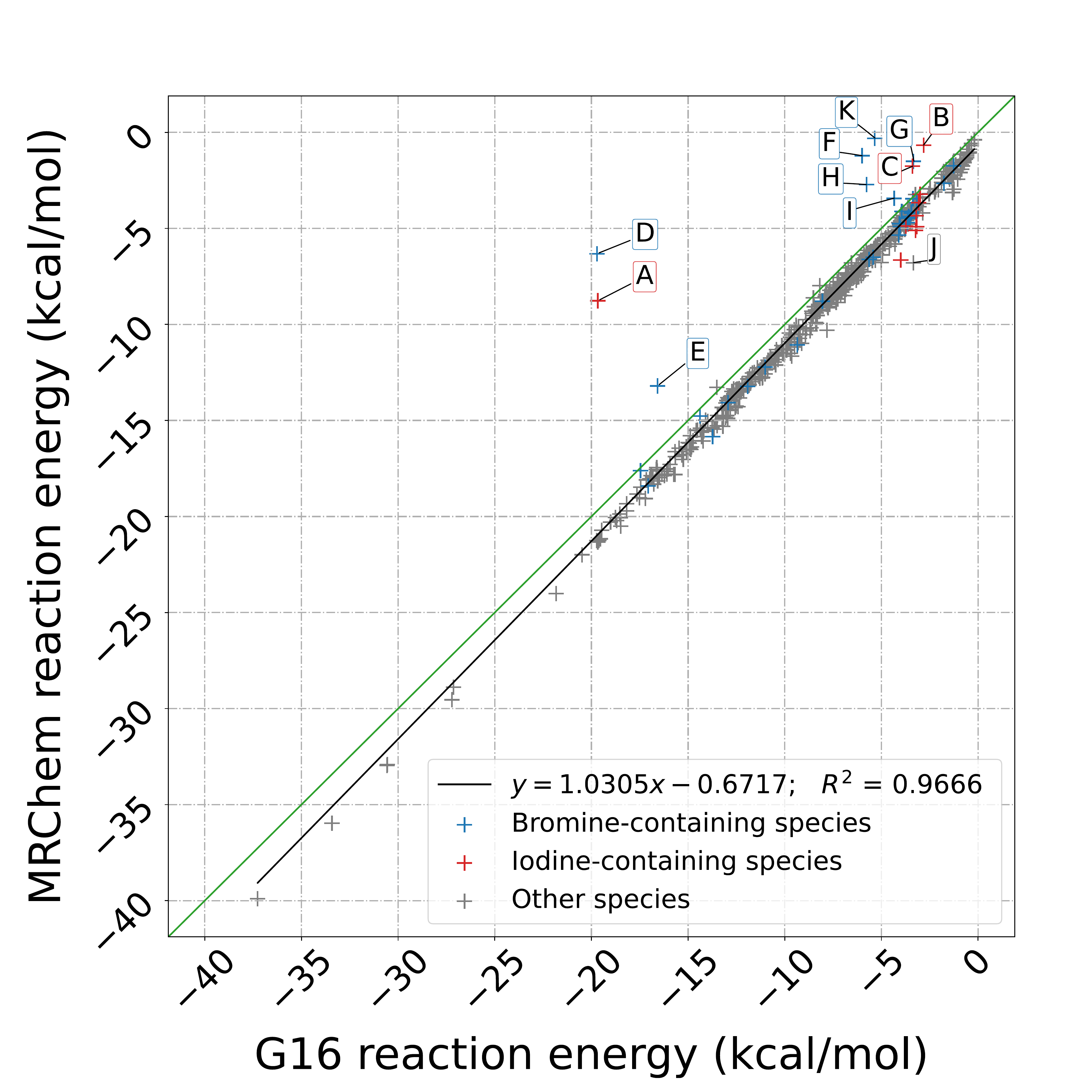}
        %\includesvg[inkscapelatex=false,width=\textwidth]{large_corr_plt_eps80_neutral.svg}
        \caption{$\varepsilon=80.0$}
        \label{fig:bigparam_a11_b5_c0_p80}
    \end{subfigure}
    \caption{Correlation plots of reaction energies computed with Gaussian16 and MRChem for all neutral species in the \ac{MSDD}\cite{Marenich2020Minnesota2012} for $\varepsilon=2.0, 4.0, 80.0$. All cavities are atom-centered, with Bondi radii.\cite{Bondi:1964fa, Mantina:2009gn} Radii are scaled by $1.1.$ in Gaussian 16. For the MRChem calculations, we used default values: $\alpha=1.1$, $\beta=0.5$, $\sigma =0.2\,\text{a.u.}$ Linear regression line shown in black.
    Outlier species are marked in blue and red when containing bromine and iodine, respectively. The labels refer to: 
A. 5-bromouracil, \ce{H3C4N2O2Br} (\texttt{n203});
B. 5-bromo-3-sec-butyl-6-methyl-uracil, \ce{H13C9N2O2Br} (\texttt{test1013}); 
C. 2-bromoanisole, \ce{H7C7OBr} (\texttt{test5008}); 
D. Bromobenzene, \ce{H5C6Br} (\texttt{0186bro});
E. 4-bromopyridine, \ce{H4C5NBr} (\texttt{0573bro});
F. 1-bromo-2-chloroethane, \ce{H4C2ClBr} (\texttt{0202bro});
G. 5-iodouracil, \ce{H3C4N2O2I} (\texttt{test2018});
H. Iodomethane, \ce{H3CI} (\texttt{test4003});
I. Iodobenzene, \ce{H5C6I} (\texttt{test4001});
J. 1,4-dichlorobenzene, \ce{H4C6Cl2} (\texttt{0176pdi});
K. 3-bromoanisole, \ce{H7C7OBr} (\texttt{test5009}).
    }
    \label{fig:bigparam-neutral}
\end{figure}

\begin{figure}[!hp]
    \centering
    \begin{subfigure}[t]{0.45\textwidth}
        \centering
        \includegraphics[width=\textwidth]{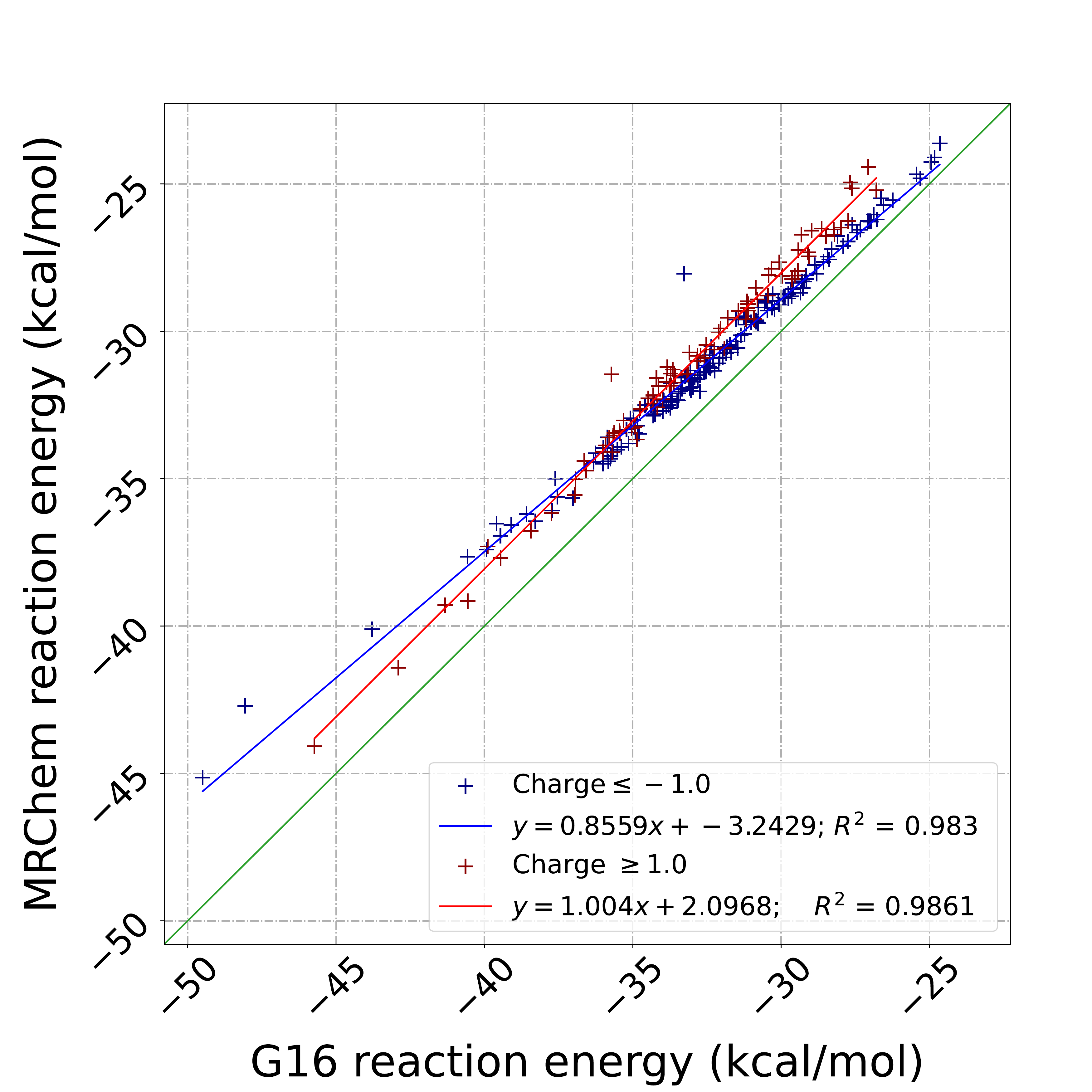}
        %\includesvg[inkscapelatex=false,width=\textwidth]{large_corr_plt_eps2_charged.svg}
        \caption{$\varepsilon=2.0$.}
        \label{fig:bigparam_a11_b5_c1_p2}
    \end{subfigure}
    \quad\quad
    \begin{subfigure}[t]{0.45\textwidth}
        \centering
        \includegraphics[width=\textwidth]{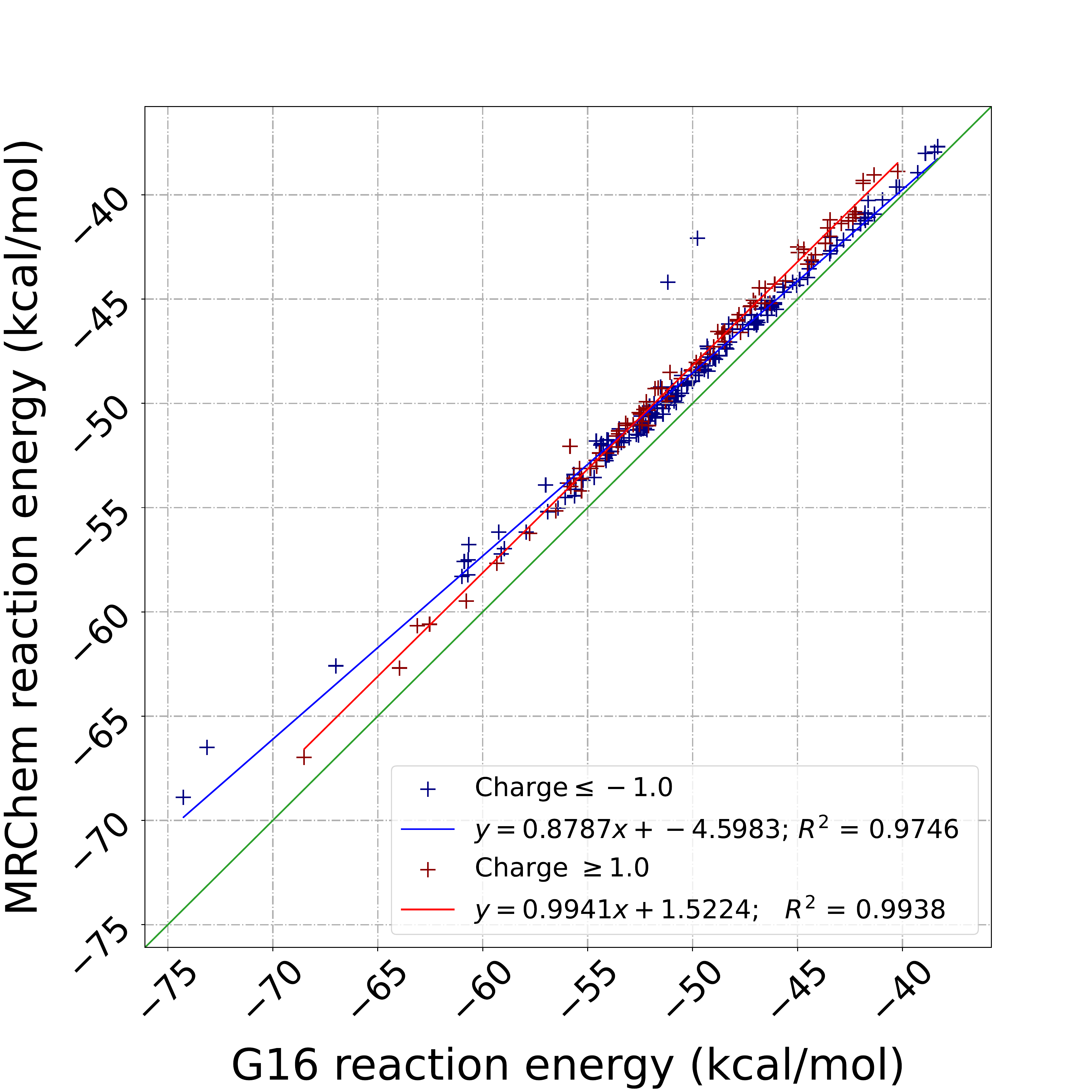}
        %\includesvg[inkscapelatex=false,width=\textwidth]{large_corr_plt_eps4_charged.svg}
        \caption{$\varepsilon=4.0$.}
        \label{fig:bigparam_a11_b5_c1_p4}
    \end{subfigure}
    \\
    \begin{subfigure}[t]{0.45\textwidth}
        \centering
        \includegraphics[width=\textwidth]{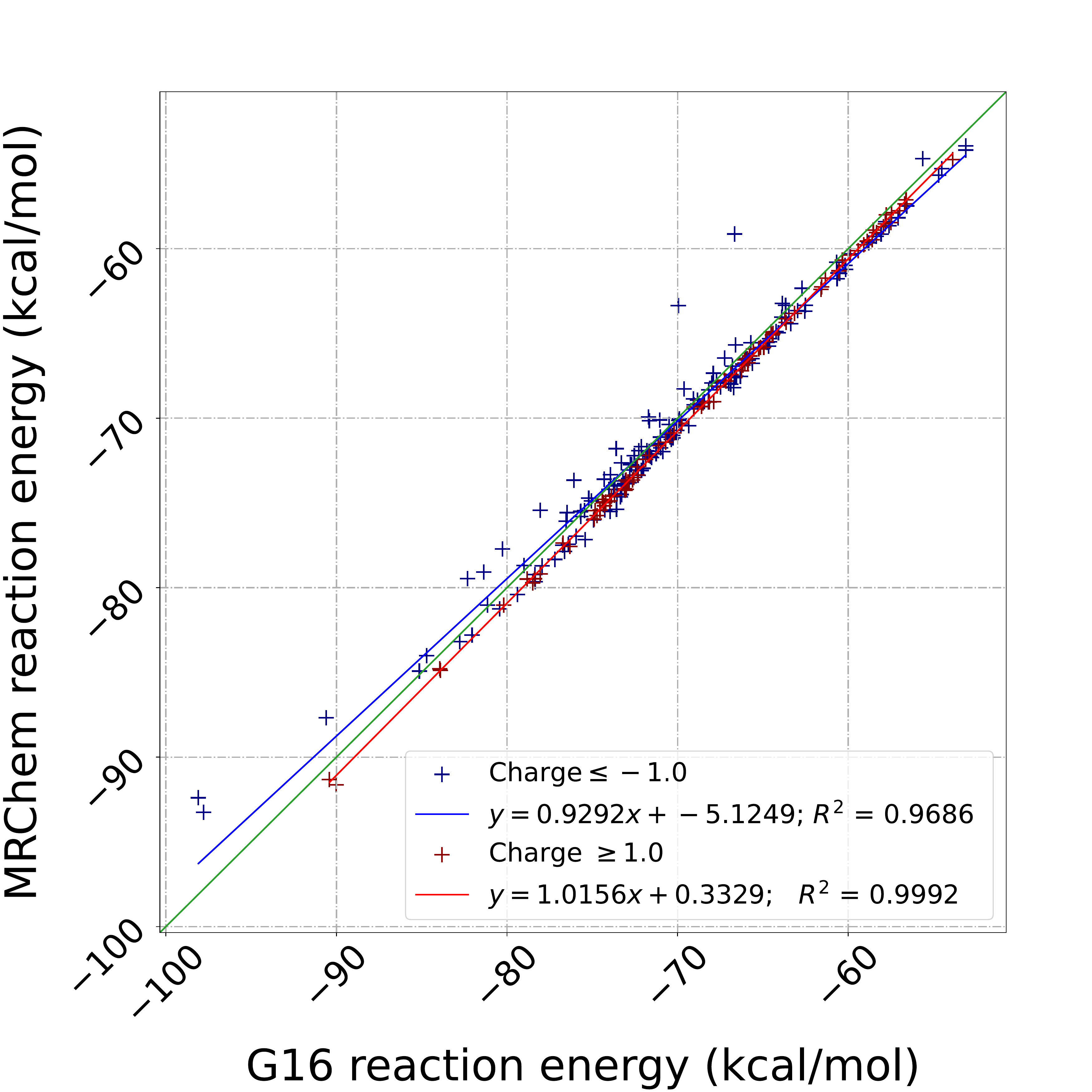}
        %\includesvg[inkscapelatex=false,width=\textwidth]{large_corr_plt_eps80_charged.svg}
        \caption{$\varepsilon=80.0$.}
        \label{fig:bigparam_a11_b5_c1_p80}
    \end{subfigure}
    \caption{Correlation plots of reaction energies computed with Gaussian16 and MRChem for all positive (red) and negative (blue) ions in the \ac{MSDD}\cite{Marenich2020Minnesota2012} for $\varepsilon=2.0, 4.0, 80.0$. All cavities are atom-centered, with Bondi radii.\cite{Bondi:1964fa, Mantina:2009gn} Radii are scaled by $1.1.$ in Gaussian 16. For the MRChem calculations, we used default values: $\alpha=1.1$, $\beta=0.5$, $\sigma =0.2\,\text{a.u.}$ Linear regression lines are shown in red (blue) for positive (negative) ions, respectively.}
    \label{fig:bigparam-charged}
\end{figure}

Figures~\ref{fig:bigparam-neutral} and~\ref{fig:bigparam-charged} summarize our results, for neutral and charged species, respectively. As for the results in Section \ref{sec:cavity-parametrization}, the figures visualize the correlation between the reaction energies computed with Gaussian16 ($x$-axis) and MRChem ($y$-axis).

For low permittivity ($\varepsilon=2.0$), Figure~\ref{fig:bigparam_a11_b5_c0_p2} 
shows that for neutral species our data is quite close to the main diagonal for small energies, but has a slight systematic deviation for more negative reaction energies (bottom left corner). For ions, Figure~\ref{fig:bigparam_a11_b5_c1_p2} shows a systematic overestimation with respect to Gaussian16, and a clear distinction between cations and anions. 
For $\varepsilon=4.0$ (Figures~\ref{fig:bigparam_a11_b5_c0_p4} and~\ref{fig:bigparam_a11_b5_c1_p4}) we see a similar trend, although most data points appear to be closer to the diagonal. 
For high permittivity $\varepsilon = 80$, Figure~\ref{fig:bigparam_a11_b5_c0_p80} for neutral species and Figure~\ref{fig:bigparam_a11_b5_c1_p80} for ionic ones, show that the values are now mostly below the diagonal, that is, solvation energies are underestimated compared to Gaussian16. 
In \ref{fig:bigparam-neutral}, we can see a set of outlying point with respect to the rest of the data. These points have been identified as species containing bromine\footnote{Molecules and corresponding filenames in the database: 
A. 5-bromouracil, \ce{H3C4N2O2Br} (\texttt{n203});
B. 5-bromo-3-sec-butyl-6-methyl-uracil, \ce{H13C9N2O2Br} (\texttt{test1013}); 
C. 2-bromoanisole, \ce{H7C7OBr} (\texttt{test5008}); 
D. Bromobenzene, \ce{H5C6Br} (\texttt{0186bro}); 
E. 4-bromopyridine, \ce{H4C5NBr} (\texttt{0573bro});
F. 1-bromo-2-chloroethane, \ce{H4C2ClBr} (\texttt{0202bro});
K. 3-bromoanisole, \ce{H7C7OBr} (\texttt{test5009}).
} or iodine,\footnote{
Molecules and corresponding filenames in the database: 
G. 5-iodouracil, \ce{H3C4N2O2I} (\texttt{test2018}); 
H. Iodomethane, \ce{H3CI} (\texttt{test4003}); 
I. Iodobenzene, \ce{H5C6I} (\texttt{test4001}).
} with only one outlier containing chlorine instead.\footnote{Molecule and corresponding filename in the database: 
J. 1,4-dichlorobenzene, \ce{H4C6Cl2} (\texttt{0176pdi}).}\cite{Marenich2020Minnesota2012}
There may be multiple, concomitant reasons for these discrepancies: (a) Bromine and Iodine are the only atoms from the fourth and fifth period of the periodic table present in the set; (b) the radii used in the definition of the cavities for these elements might not be appropriate; (c) the different treatment of volume polarization in the two implementations (full account in our model and implicit first-order correction in the \ac{IEFPCM} model\cite{Chipman:2000eh, Cances:2001eh}) might affect the description of these molecules, where a more delocalized electronic density is expected. It would be interesting to disentangle the effects of surface and volume polarization, but it is not straightforward to do so and it goes beyond the scope of the present work.

 %For a fixed permittivity, the molecules in the database have a much smaller reaction energy than the ions, as shown by comparing figures \ref{fig:bigparam-neutral} and \ref{fig:bigparam-charged} for the same permittivity values. The correlation for low permittivity values formolecules is better when the reaction energy is closer to zero. For $\epsilon = 80.0$ there seems to be a smaller improvement than for the lower permittivity values.
 
The fact that molecules are quite close to the line, especially as the reaction energy becomes small (top right corner), is not surprising. We chose the cavity parameters from a limited set of small molecules. On the other hand, the observed deviations for larger solvation energy are to a large degree systematic, which shows that they could be accounted for, with a more refined parametrization.

Cations tend to have less diffuse density than anions. Therefore, the size of the cavity with respect to the spatial extent of the electronic density is larger for cations than for anions. According to the simple Born model, solvation energy of ions is inversely proportional to the radius of the cavity, which explains the better correlation observed for cations: when the charge distribution is better confined inside the cavity, the difference between a sharp interface formally not accounting for volume polarization and a diffuse one including it, becomes smaller.

\subsection{Performance}

The current code is a prototype and we have therefore not yet dedicated attention to improving its performance in terms of computational time and memory footprint. A few general considerations can however be made. The solution of the \ac{GPE} is technically similar to that of the Helmholtz equation which we employ to solve the \ac{SCF} equations.\cite{Jensen2022-wd,Harrison:2004vua} It should therefore be possible to achieve linear scaling with respect to the system size once the code is fully optimized.\cite{Wind2022-cg} This is a feature of \ac{MRA},\cite{Alpert2002-sr} which is designed to decouple the long- and short-range interactions automatically thanks to the adaptive refinement scheme coupled with the use of the non-standard form of operators.\cite{Beylkin:2008im}
In this sense, the algorithm should be competitive with implementations of sharp-cavity models that employ the \ac{FMM} to accelerate the matrix-free solution of the \ac{PCM} equations.\cite{Scalmani2004-bp}

A qualitative comparison with the \ac{DD} family of algorithms\cite{Cances:2001eh,Stamm2016-fv,Stamm2019-tr} is also in order. \ac{DD} approaches to implicit solvation are, by construction, linear scaling. Furthermore, they are easily recast in matrix-free form which both reduces the memory footprint and lends itself to further performance boosting \emph{via} the \ac{FMM}.\cite{Mikhalev2022-gq}
However, in our understanding of the algorithm, these advantages of the method are not straightforwardly extended to cavities with diffuse boundaries. Furthermore, when dealing with quantum mechanical source densities, the quantum-classical coupling must rely on volume integrations, \emph{e.g.} using a DFT grid, to correctly represent the escaped charge.\cite{Nottoli2019-sn}

Our algorithm achieves formal simplicity and, in principle, algorithmic efficiency. Real-space methods for the reaction potential can be coupled with \ac{GTO} methods for the electronic-structure problem,\cite{Howard2017-sa} thus making our method of interest \emph{beyond} multiwavelet-based quantum chemistry.
Currently the main bottleneck is constituted by the memory footprint of the functions describing the cavity and the solvent reaction potential, since they extend throughout the whole computational domain. Work is currently in progress to deal with such functions in an efficient way.

\section{Conclusions}\label{sec:conclusions}

We have implemented, parametrized and benchmarked a continuum solvation model based on a position dependent permittivity $\varepsilon(\vect{r})$.\cite{FossoTande:2013ka}
Our algorithm performs microiterations, nested within each \ac{SCF} cycle, to obtain the solvent reaction potential.
We overcome convergence issues using \ac{KAIN} convergence acceleration and an adaptive convergence threshold. Our implementation is robust and introduces only a modest computational overhead.

With a simple parametrization, we have obtained a good correlation with respect to the \ac{IEFPCM} implemented in Gaussian16, for an extensive library of geometries and a wide range of permittivities. Some systematic deviations have been observed, suggesting that a more sophisticated cavity parametrization could yield even better agreement.
An alternative option, which is often challenging for standard solvation models, is to parametrize the permittivity by making use of an isodensity cavity as support. This choice would forego the radius parametrization altogether, but it might pose other challenges, because the cavity gradient must be computed numerically, and the coupling with the density functional must be taken into account.

The performance and stability might be further improved, by considering a different approach to the \ac{SCRF} microiterations: a square-root parametrization of the electrostatic potential, as suggested by \citeauthor{Fisicaro:2016kl}, might prove useful.\cite{Fisicaro:2016kl}

The flexibility of the method will allow for several additional developments, such as the inclusion of charged particles outside the cavity, as well as other contributions to the solvation energy, such as cavitation, dispersion and repulsion.

\section{Code and data availability}

Input and output files for the Gaussian16 and MRChem calculations reported in this work are available
on the Norwegian instance of the Dataverse data repository: \url{https://doi.org/10.18710/TFSWLC}. The data package also includes the Jupyter notebooks used to produce the graphs in this paper.

\section{Author contributions}
\label{author_contrib}

We use the CRediT taxonomy of contributor roles.\cite{Allen:2014kj,Brand:2015jr}
The ``Investigation'' role also includes the ``Methodology'', ``Software'', and ``Validation'' roles.
The ``Analysis'' role also includes the ``Formal analysis'' and ``Visualization''
roles. The ``Funding acquisition'' role also includes the ``Resources'' role.
We visualize contributor roles in the following authorship attribution matrix, as suggested in Ref.~\citenum{authorship-attribution}.

\begin{table}
  \caption{Levels of contribution: \textcolor{blue!100}{major}, \textcolor{blue!25}{support}.}
  \label{tbl:example}
  \begin{tabular}{lccccc}
    \toprule
                              & GAGS   & RDRE   & SRJ    & MB     & LF     \\
    \midrule
    Conceptualization         &        & \major &        &        & \major \\ 
    Investigation             & \major & \minor & \minor & \minor & \minor \\ 
    Data curation             & \major &        &        &        &        \\ 
    Analysis                  & \major & \minor & \minor &        & \minor \\ 
    Supervision               &        & \minor & \minor & \minor & \major \\ 
    Writing -- original draft & \major & \major & \minor &        & \minor \\
    Writing -- revisions      & \major & \major & \minor & \minor & \major \\
    Funding acquisition       &        &        &        &        & \major \\ 
    Project administration    & \minor &        &        &        & \major \\
    \bottomrule
  \end{tabular}
\end{table}

\begin{acknowledgement}
We acknowledge support from the Research Council of Norway through its Centres
of Excellence scheme, project number 262695, and through the FRIPRO grant ReMRChem (324590), from the Tromsø Research Foundation (SURFINT, A32543) and from Notur -- The Norwegian Metacenter for Computational Science through grant of computer time, no.
nn4654k. R.D.R.E. acknowledges support from the European High-Performance Computing  Joint Undertaking under Grant Agreement No. 951732 and partial support by the Research Council of Norway through its
Mobility Grant scheme, project number 261873.
We thank Simone Brugiapaglia (Concordia University) for helpful discussions on a point raised by one of the reviewers.
\end{acknowledgement}

\appendix

\section{Analytical derivatives of the permittivity and cavity functions}\label{app:eps-and-C-derivatives}
\subsection{The gradient}
The gradient of the permittivity function can be determined analytically. Differentiating Equation~\ref{eq:expperm}:
\begin{equation}
    \begin{aligned}
   \nabla\permittivity 
   &=  
   - \varepsilon_{\mathrm{in}} 
   \exp\left[
   \left(1 - C{\left(\mathbf{r}\right)}\right) 
   \log{\left(\frac{\varepsilon_{\mathrm{out}}}{\varepsilon_{\mathrm{in}}} \right)}
   \right] 
   \log{\left(\frac{\varepsilon_{\mathrm{out}}}{\varepsilon_{\mathrm{in}}} \right)} \nabla C{\left(\mathbf{r}\right)} 
   \\
   &= 
   \log{\left(\frac{\varepsilon_{\mathrm{out}}}{\varepsilon_{\mathrm{in}}} \right)}
   \permittivity 
   \nabla C{\left(\mathbf{r}\right)},
    \end{aligned}
\end{equation}
which only requires to compute the analytical gradient of the interlocking sphere cavity function $C(\vect{r})$.

The analytical gradient of the interlocking sphere cavity is as defined by \citeauthor{FossoTande:2013ka}:\cite{FossoTande:2013ka}
\begin{equation}
    \nabla C (\mathbf{r})
    = 
    \left[\prod_{\alpha=1}^{N_{\mathrm{sph}}} 
    \left(1 - {C_{\alpha}}{\left(\mathbf{r} \right)}\right)\right]
    \sum_{\alpha=1}^{N_{\mathrm{sph}}} 
    \frac{\nabla {C_{\alpha}}{\left(\mathbf{r} \right)}}{1 - {C_{\alpha}}{\left(\mathbf{r} \right)}}.
\end{equation}
The gradient of a single sphere cavity function $C_{\alpha}$ is:
\begin{align}
    \nabla {C_{\alpha}}{\left(\mathbf{r} \right)} 
    = 
    - 
    \frac{1}{\sigma \sqrt{\pi}}
    \exp\left(
    \frac{{s_{\alpha}}^{2}{\left(\mathbf{r} \right)}}{\sigma^{2}}
    \right) 
    \nabla {s_{\alpha}}(\mathbf{r} ),
\end{align}
and finally the gradient of the signed normal distance is:
\begin{align}
    \nabla s_{\alpha}(\mathbf{r}) = 
    \begin{pmatrix}
    \frac{x - x_{\alpha}}{\left|\mathbf{r} - \mathbf{r_{\alpha}}\right|} \\
    \frac{y - y_{\alpha}}{\left|\mathbf{r} - \mathbf{r_{\alpha}}\right|} \\
    \frac{z - z_{\alpha}}{\left|\mathbf{r} - \mathbf{r_{\alpha}}\right|}
    \end{pmatrix}
\end{align}
In the implementation, we use a cutoff of $10^{-12}$ for the denominator, in order to avoid numerical discontinuities.

\bibliography{references} 
%\bibliography{references,PCMMW_refs_new} 

\end{document}